\documentclass[9pt,twocolumn,twoside]{osajnl}
\usepackage{multirow}
\usepackage{tabularx}
\usepackage[normalem]{ulem}
\journal{jocn} 

\setboolean{shortarticle}{false}

\title{Toward Agentic Optical Networks: A Vision of LLM Agent-Driven Autonomous Lifecycle Management}

\author[1,+]{Yao Zhang}
\author[1,+]{Shengnan Li}
\author[1]{Yuchen Song}
\author[1]{Yidi Wang}
\author[2]{Yue Pang}
\author[1]{Wenbin Chen}
\author[3]{Xiaotian Jiang}
\author[1]{Xiao Luo}
\author[4]{Meixia Fu}
\author[1]{Min Zhang}
\author[1]{Yongli Zhao}
\author[1]{Shanguo Huang}
\author[5]{Alan Pak Tao Lau}
\author[1,*]{Danshi Wang}

\affil[1]{State Key Laboratory of Information Photonic and Optical Communication, Beijing University of Posts and Telecommunications, Beijing, 100876, China}
\affil[2]{China Telecom Cloud Network Operating System R\&D Center, Beijing, 102299, China}
\affil[3]{State Key Laboratory of Optical Fiber and Cable Manufacture Technology, China Telecom Research Institute, Beijing, China}
\affil[4]{School of Automation and Electrical Engineering, Institute of Industrial Internet, University of Science and Technology Beijing, Beijing, 100083, China}
\affil[5]{Photonics Research Institute,
Department of Electrical Engineering, The Hong Kong Polytechnic University, Kowloon, Hong Kong, SAR, China}

\affil[+]{These authors contributed equally to this work.}

\affil[*]{danshi\_wang@bupt.edu.cn}

\begin{abstract}
As optical networks continue to expand in scale, complexity, and service diversity, the implementation of automation has become essential for ensuring agility, efficiency, and reliability in lifecycle management (LCM) of optical networks. Large language model (LLM) Agent, distinguished by its progressively sophisticated capabilities in logical reasoning, adaptive decision-making, complex problem solving, and multi-task orchestration, presents great opportunities to advance network automation beyond traditional AI techniques. Nevertheless, the application of LLM Agent in optical networks remains in its early exploratory stage, challenged by the lack of multi-task coordination, high computational demands, data dependence, and reliability concerns. In this paper, we envision a conceptual roadmap toward Agentic Optical Networks (AONs) by integrating LLM Agents throughout the LCM with high-level autonomy. First, we trace the evolution from manual operations to AI-empowered frameworks and distill key technologies in Agent, providing actionable insights into leveraging its strengths for addressing practical network automation challenges. A core contribution of this paper is the proposal of a hierarchical multi-Agent framework, which is specifically developed to manage every phase in LCM of AONs, including planning, deployment, operation, maintenance, upgrade, and decommission, thereby enabling more cohesive and comprehensive automation throughout the entire lifecycle. In addition, future directions and underlying challenges are also discussed at the intersection of LLM and optical networks. By aligning the LLM Agent with the specialized requirements of AONs, this work aims to explore the potential for the evolution of optical networks moving from task-level semi-automatic execution toward lifecycle-level full autonomy.
\end{abstract}

\setboolean{displaycopyright}{false} 

\begin{document}

\maketitle

\section{Introduction}
The lifecycle management (LCM) of optical networks represents a systematic framework to govern the entire lifespan of optical infrastructure, involving six critical phases: planning, deployment, and operation, maintenance, upgrade, and decommissioning \cite{song2025lifecycle}. Serving as the backbone of modern telecommunications, optical networks are engineered as large-scale complex systems, consisting of cross-connected optical fibers, expanding nodes, and massive optical components. Traditional LCM methodologies, constrained by manual interventions and reactive operations, struggle to address complex LCM of modern optical networks, which requires proactive and intelligent governance \cite{lam2024quantifying}. Propelled by the breakthroughs in artificial intelligence (AI), the paradigm is currently evolving toward autonomous LCM \cite{wang2022review} characterized by closed-loop autonomy.

Throughout the evolution of optical networks, AI techniques have progressively contributed distinct capabilities across different levels of autonomy. Since the 2010s, basic machine learning (ML) methods have facilitated data-driven and efficient solutions for tasks primarily involving classification and clustering \cite{gu2020machine}. Meanwhile, advanced deep learning (DL) techniques have gained significant attention in optical networks \cite{musumeci2018overview, lu2021performance} owing to their powerful learning capabilities, enabling the high-precision modeling of complex nonlinear relationships by mapping input data to desired outputs \cite{wang2021artificial}. However, both ML and DL models are typically tailored for specific tasks and exhibit limited generalization across diverse problem domains. Specifically, it is necessary to retrain or fine-tune separate models for different applications, such as performance prediction, parameter identification, and configuration optimization. Furthermore, these models generally lack the ability to comprehend or analyze the broader network context, which is an essential cognitive function required for adaptive decision-making \cite{sui2020review}. As a result, these techniques facilitate foundational levels of automation in optical networks but fall short of supporting high-level automation. Advancing toward fully autonomous optical networks necessitates the integration of more advanced AI techniques with general intelligence, thereby enabling models to understand, reason, and make decisions based on a comprehensive view of network state and operational intent.

In recent years, generative AI (GenAI) has experienced rapid advancements \cite{cao2023comprehensive} marked by the emergence of large language models (LLMs) that exhibit remarkable capabilities in generating coherent text, image, code, and even structured solutions. With powerful reasoning and cognitive capabilities, LLM can handle a wide range of tasks through prompt-based interaction and in-context learning, thereby creating innovative opportunities for AI applications in specialized domains \cite{bariah2024large}. In optical networks, initial efforts have explored LLM Agents to solve multiple tasks \cite{wang2024large, song2025synergistic, sun2025experimental, liu2025first}, including performance optimization, fault diagnosis, resource orchestration, and device control. By leveraging their reasoning and cognitive capabilities, LLM Agents aim to overcome the fragmentation and scalability limitations of conventional AI solutions, thereby facilitating more integrated, autonomous, and adaptive capabilities for future Agentic Optical Networks (AONs).

\begin{figure*}[ht!] 
\centering
\includegraphics[width=14cm]{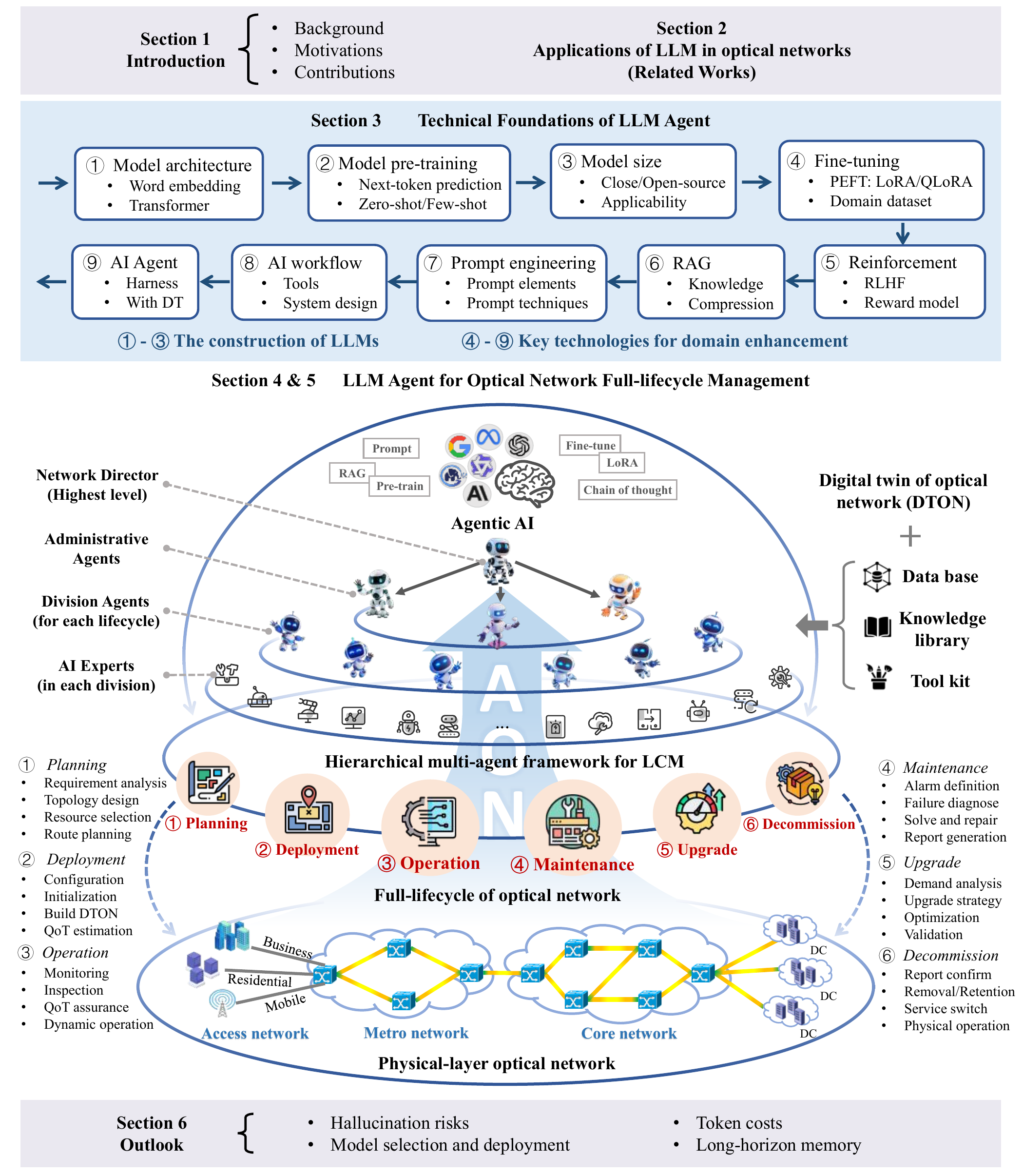}
\caption{A comprehensive overview of the structure and key topics covered in this work.}
\label{fig-1}
\end{figure*}

Predictably, building upon LLM capabilities, the paradigm of optical network is evolving toward Agent-based systems. Nevertheless, the application of LLM Agent in optical networks is still in its early exploratory stage. Several key challenges remain to be addressed before its potential can be fully harnessed across the LCM process \cite{wang2025OFC, zhang2026ai}. First, the deployment of Agents faces practical constraints, including the high computational cost, large model sizes, and substantial token consumption, which limit their scalability and adaptability in real-world scenarios. Second, while techniques such as prompt engineering, context engineering, and harness engineering have shown promise in adapting to specific tasks, a universally efficient methodology for optical network's LCM has yet to be established. Third, given the diverse and complex tasks throughout the entire lifecycle of optical networks, it is difficult for a single monolithic model to fully meet all functional and operational requirements. Therefore, there is a pressing need for a multi-agent collaborative framework to support modular deployment and facilitates coordinated execution across diverse tasks. Furthermore, addressing issues such as hallucination and reliability is essential to ensure safe and trustworthy automation in network operations. These challenges highlight the necessity of developing Agentic frameworks and elaborating collaborative workflows that are specifically tailored to the LCM of AONs.

In this paper, we present a comprehensive roadmap for integrating LLM Agent into the full lifecycle of AONs. The main contributions are as follows: 1) From a network-centric perspective, we review the application of LLMs in optical networks and summarize representative related works, highlighting the potential of LLM Agents to enhance reasoning, decision-making, and automation capabilities. 2) We introduce several core technologies in LLM-driven Agent development, providing the optical network community with guidance on leveraging these capabilities for automation tasks. 3) As the main contribution, a hierarchical multi-Agent framework is proposed to implement the autonomous LCM of AONs, including network planning, deployment, operation, maintenance, upgrade, and decommission phases. This organization extends Agent-based automation from individual operational and management tasks toward the full network lifecycle, while providing clear task ownership and modular scalability as network functions and lifecycle requirements evolve. 4) Finally, several key challenges at the intersection of LLM Agents and optical networks are outlined. By systematically aligning LLM technologies with the specific demands of AONs, we provide a vision of Agent-enhanced automation throughout the entire lifecycle, paving the way toward fully self-driving optical networks that will underpin the next generation of intelligent connectivity.

\section{Applications of LLMs in Optical Networks}

\begin{table*}[t]
\centering
\caption{Representative works integrating LLMs with optical networks}
\label{table1}
\renewcommand{\arraystretch}{1.5}
\setlength{\tabcolsep}{4pt}
\footnotesize
\begin{tabular}{p{3cm} p{9cm} p{3cm} p{1.5cm}}
\hline
\textbf{Applications} &
\textbf{Capabilities and scope} &
\textbf{Method Type} &
\textbf{Ref} \\
\hline

\multirow{2}[4]{=}{Framework of LLM-driven optical network} 
& LLM applications, challenges, and opportunities for intelligent network operations, prediction, and analysis 
& GPT-based prompt & \cite{wang2024large} \\ \cline{2-4}
& Capabilities and limitations of applying LLMs to optical networks
& Conceptual & \cite{wang2025OFC, cruzes2024revolutionizing, wang2025GenAIworkshop, zhang2026ai} \\ \hline

\multirow{2}[1]{=}{Alarm/Log analysis }
& \multirow{2}[1]{=}{LLM-enabled professional alarm/log Q\&A, analysis, and reasoning}
& GPT-based prompt & \cite{wang2024alarmgpt} \\ \cline{3-4}
& & LLaMA fine-tune & \cite{pang2024large} \\ \hline

QoT estimation
& LLM-assisted QoT estimation and DT-based multi-task network management
& GPT-based prompt & \cite{zhang2024gpt} \\ \hline

Failure management
& LLM-based fault analysis and management fused multi-mode data
& GPT-based prompt & \cite{yang2024spatio} \\ \hline

\multirow{2}[1]{=}{Configuration and control }
& \multirow{2}[1]{=}{Practical pipeline for automated configuration in SDN network and testbeds }
& Prompt & \cite{Cicco2024open, zhou2024LLM} \\ \cline{3-4}
& & Qwen fine-tune & \cite{wang2025LLMcentric} \\ \hline

Simulation assistant
& System simulation and performance evaluation, with natural language
& GPT-based prompt & \cite{jiang2024opticomm} \\ \hline

Network operations automation
& LLM-assisted network operation workflows with task automation and tool interaction
& LLaMA fine-tune & \cite{sun2024PDP, Allen2025endtoend, sun2025experimental} \\ \hline

Field-trial demonstration
& Field-trials for LLM-assisted lifecycle management with interaction with network management and operational tools
& Prompt, RAG & \cite{song2025synergistic, liu2025first, zhang2025first, huang2025field} \\ \hline

Performance optimization
& Integrates human expertise to enhance efficiency and reliability in AI Agent
& AI Agent, GPT-4o & \cite{qiu2025expertise, zhang2025design} \\ \hline

Autonomous multi-task collaboration
& Multi-Agent collaboration for task decomposition, coordination, and tool calling
& Multi-Agent framework & \cite{hao2025multi, zhang2025generative, xiang2025optima} \\ \hline

\end{tabular}
\end{table*}

Currently, LLMs, as a representative form of GenAI, mark a significant leap forward in natural language processing (NLP), transforming how the language is understood and generated. The rapid advancement of LLM technologies also presents promising opportunities to automate numerous tasks in the field of optical networks \cite{wang2024large}. Recent efforts to integrate LLMs into optical networks, as illustrated in Table \ref{table1}, can be broadly categorized into three progressive stages.

In the initial phase, research focused on leveraging LLMs' capabilities in context understanding and generation to assist in basic log and alarm processing tasks. For instance, AlarmGPT \cite{wang2024alarmgpt} and instruction-tuned LLaMA \cite{pang2024large} demonstrated how LLMs could automatically summarize alarms, parse logs, and generate structured outputs, thereby reducing reliance on manual intervention. These approaches primarily treated LLM as intelligent text processors embedded in optical network toolchains. As exploration deepened, researchers began incorporating domain-specific knowledge into prompts and utilizing LLMs for workflow decomposition and task chaining. For example, in quality of transmission (QoT) estimation and network planning tasks, LLMs were employed to involve physical models, analyze simulation results, suggest parameter values, in a loop-like manner \cite{zhang2024gpt, jiang2024opticomm}. Fault management tasks also benefited from LLMs’ ability to reason across symptoms and suggest candidate root causes \cite{yang2024spatio}. 

More recently, with the growing demand for autonomous operation in optical networks, LLMs have begun to take on Agent-like roles. Instead of merely assisting with textual processing or isolated task reasoning, they are increasingly integrated into broader network systems, enabling interaction with real-time data, tools, and control environments \cite{song2025synergistic}. One direction involves coupling LLMs with digital twin (DT) platforms, allowing them to perceive dynamic network states and participate in closed-loop decision-making. For instance, Sun et al. \cite{sun2025experimental} explored an AI Agent for optical network management, where the LLM interprets simulation outputs and responses operational adjustments. Similarly, Liu et al. \cite{liu2025first} demonstrated how LLM-powered Agent could continuously monitor network and adapt responses. These efforts underscore a transition from data interpretation to proactive execution and feedback for complex tasks. 

Another emerging trend is the integration of LLMs with network control and management interfaces, such as software-defined network (SDN) controllers or network management system (NMS). This includes efforts to map user intents to structured network commands via programmable interfaces \cite{Ricard2024applying, wang2025LLMcentric}, as well as architectures that integrate LLMs directly into service orchestration pipelines and configuration workflows \cite{Cicco2024open, zhou2024LLM}. Further developments have demonstrated frameworks for end-to-end task execution \cite{Allen2025endtoend}, where LLM Agent not only interpret operational goals but also coordinate resource allocation and execute network updates in response to environmental changes. Building upon these single-Agent foundations, recent breakthroughs have increasingly shifted towards deploying locally-hosted, open-source LLMs to address data privacy and latency concerns in real-world scenarios. For example, field trials have demonstrated AI Agents powered by local open-source models achieving full LCM in elastic optical networks, executing autonomous service provisioning and failure recovery within minutes \cite{huang2025field}. Moreover, structured human expertise has been explicitly integrated into the prompts \cite{qiu2025expertise,zhang2025design} to enhance domain adaptability and reliability of these Agents.

As network scenarios become more complex, the paradigm is rapidly evolving from single-Agent assistants to collaborative multi-Agent systems. In these architectures, complex workflows are decomposed and assigned to specialized Agents that communicate and collaborate to prevent the accumulation of errors \cite{zhang2025generative}. Similarly, multi-Agent frameworks have been applied to specific component-level tuning, such as autonomously control Raman amplifiers \cite{xiang2025optima}, or for strategy generation with DT interaction \cite{hao2025multi}. These multi-Agent strategies, closely aligned with the vision of fully autonomous, intent-driven networks \cite{OFC2026harish, OFC2026auto}, represent a transformative leap toward self-operating infrastructures.

Overall, these explorations highlight the feasibility of applying LLMs to various network tasks, and showcase the potential of Agent to enhance efficiency, reduce human workload, and support decision-making in complex, dynamic optical systems. From the perspective of lifecycle coverage, existing studies span individual network functions, selected operational stages, and increasingly multiple lifecycle phases, including recent field-trial demonstrations. In terms of Agent organization and task orchestration, existing approaches range from single LLM-based assistants and task-specific Agents to multi-Agent schemes for task decomposition, coordination, and collaborative execution. Regarding system interaction, prior studies explored integration of LLMs and Agents with DTs, NMSs, and domain-specific tools to different extents. These efforts should be viewed as complementary steps in the evolution of LLM-enabled optical network automation, collectively illustrating a progression from task-level assistance toward increasingly coordinated and autonomous network management.

From this perspective, rather than replacing existing task-specific or workflow-oriented approaches, this work adopts a lifecycle-oriented perspective and organizes Agent capabilities around the major phases of an optical network. We aim to synthesize these developments from a system-level architectural perspective and to further explore how heterogeneous Agent capabilities can be organized with network management systems and domain-specific tools to progress from task-level LLM assistance toward full-lifecycle autonomous management. Moving forward, further improvements can be achieved by leveraging advanced techniques such as instruction tuning, reinforcement learning (RL), fine-tuning, tool augmentation, retrieval-augmented generation (RAG), and harness engineering, all of which are discussed in the following sections. These techniques offer the promise of building more robust, reliable, and domain-aligned LLM Agents tailored to the unique demands of optical networks.

\section{Technical Foundations of LLM Agent}

Building upon the recent advances in LLMs, this section delves into the underlying technical foundations that equip LLM to be Agent to address various complex tasks. We focus on the core architectures, particularly transformer, as well as the mechanisms that enable LLMs to generalize across domains. This technical overview provides the necessary background for understanding how LLMs can be adapted and applied to autonomous optical network management in subsequent sections.

\subsection{The construction of LLM}

Following the seminal work by Vaswani et al. \cite{vaswani2017attention}, Transformer-based models have become central to modern NLP and beyond due to their scalable and context-aware modeling capability. In NLP, input text is first divided into discrete units called tokens, which may correspond to words, subwords, or characters depending on the tokenizer. Each token is then mapped to a dense numerical vector through an embedding layer, providing a continuous representation that can be processed by the neural network. The core innovation of Transformer, self-attention, enables each token to dynamically evaluate the relevance of other tokens in the same input sequence and aggregate their information into a context-aware representation. While multi-head attention performs multiple attention operations in parallel, allowing the model to capture different semantic and structural patterns simultaneously \cite{bahdanau2014neural, dosovitskiy2020image}. The performance and reliability of LLMs are fundamentally shaped by the data trained on, which must balance linguistic diversity and domain relevance \cite{pre-data1}. Large-scale general text sources support broad language understanding, while domain-specific resources, such as optical network logs, standards, protocols, and technical literature, improve reasoning in specialized contexts \cite{zhao2023survey}. Data preprocessing, including quality filtering, deduplication, and sensitive information removal, further enhances representation quality and model alignment with practical usage \cite{pre-data4, pre-data7}.

\begin{figure*}[t]
\centering
\includegraphics[width=16cm]{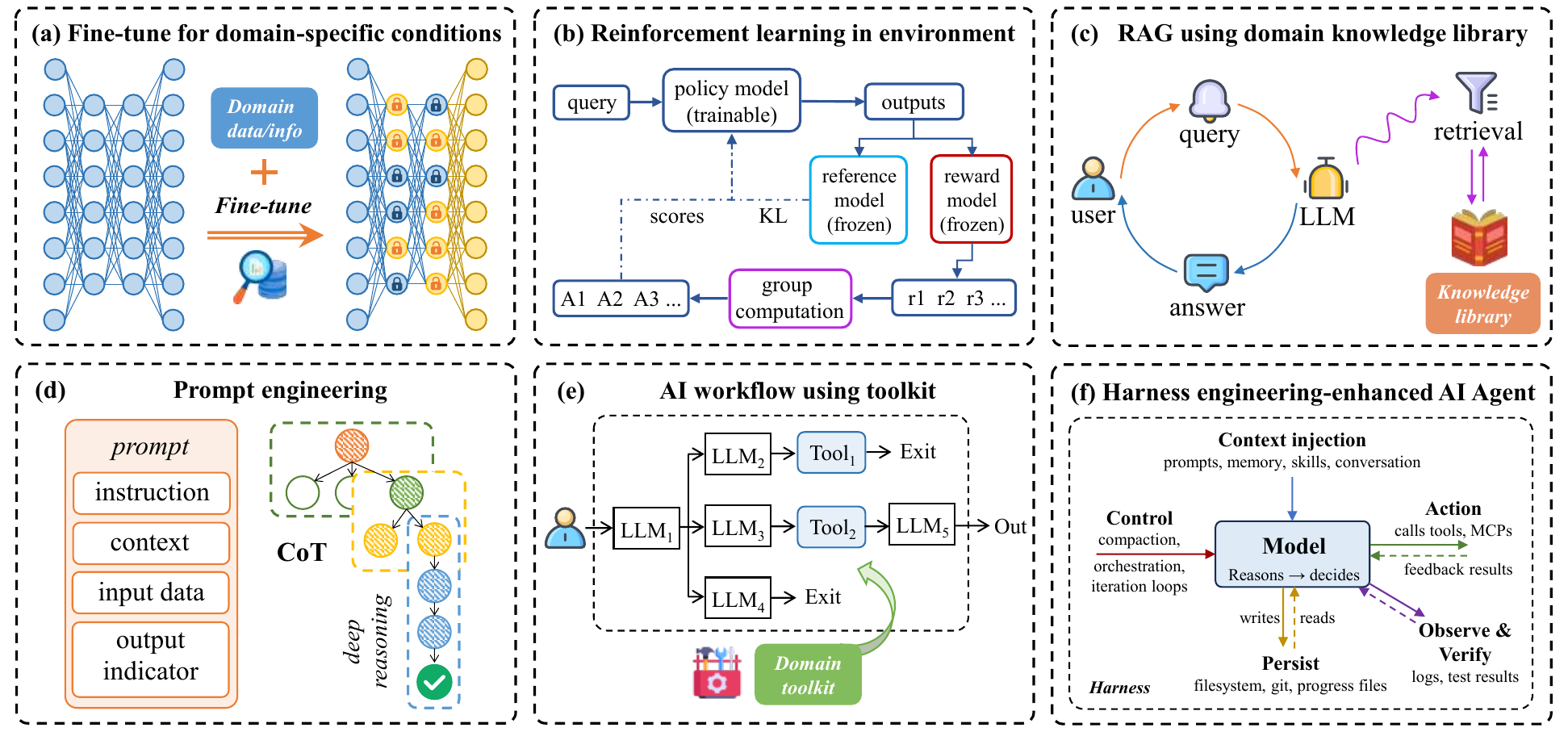}
\caption{Key technologies for domain enhancement. (a) Fine-tune (b) Reinforcement learning (c) RAG (d) Prompt engineer (e) AI workflow (f) Harness engineering-enhanced AI Agents.}
\label{fig-3.C}
\end{figure*}

Pre-training enables LLMs to learn linguistic structure and reasoning patterns from large unlabeled corpora. Most state-of-the-art LLMs, including GPT \cite{pre-train1} and LLaMA \cite{pre-train2}, adopt autoregressive decoder-only architectures that predict tokens sequentially. As representations propagate through successive layers, they accumulate increasingly contextual information \cite{devlin2018bert}. Importantly, well-pretrained models often exhibit strong zero-shot and few-shot learning abilities \cite{radford2019language}. In zero-shot learning, the model performs a task without task-specific examples being provided in the prompt, relying on it pretrained knowledge and the task instruction. In few-shot learning, a small number of task examples are provided in the prompt to guide the model toward the desired behavior, without updating its parameters. These capabilities allow a single pretrained model to adapt to different tasks through instructions or examples, providing an important foundation for its application to heterogeneous optical-network management tasks.

Current pretrained LLMs can be broadly categorized into closed-source and open-source models. Closed-source models such as OpenAI’s GPT, Google’s Gemini, and Anthropic's Claude, are characterized by extremely large training corpora and parameter scales, providing state-of-the-art general and even domain-specific reasoning capabilities, including advanced cross-modal understanding and generation. They are simple to use and highly effective for solving complex problems. In optical network scenarios, such models are therefore more suitable for non-sensitive, high-level tasks, for instance, conceptual architecture design, early-stage algorithm prototyping, or exploratory analysis of network behaviors when real operational data is not involved. Open-source models such as Qwen, DeepSeek, LLaMA, and GLM, support domain customization through fine-tuning and reinforcement learning. These models are particularly suitable for optical network scenarios that require customization and strict data governance, such as building performance estimation assistants, generating device-level configurations, supporting troubleshooting assistants, or enabling closed-loop control modules that must run within operator environments with proprietary data.

\subsection{Key technologies of LLM for vertical domain adaptation}

\emph{1) Fine-tuning for domain task enhancement}

While pre-trained foundation models exhibit strong general reasoning capabilities, they may lack the specialized knowledge required for optical networks, where engineering decisions depend on field monitoring performances, physical-layer constraints, and device-specific standards, that rarely appear in general pretraining corpora. Therefore, fine-tuning becomes the key mechanism to bridge this gap, by further training models in domain-specific datasets, enabling models to internalize optical network concepts, operational patterns, and engineering heuristics \cite{finetuning1}, as illustrated in Fig. \ref{fig-3.C} (a). Several fine-tuning strategies support domain adaptation with different computational and data requirements. Full-parameter fine-tuning updates the entire model but is resource intensive, whereas partial and parameter-efficient fine-tuning (PEFT) methods \cite{finetuning2} offer more practical solutions when labeled data and computational resources are limited. Techniques such as low-rank adaptation (LoRA) and Quantized LoRA (QLoRA) reduce training overhead while maintaining strong adaptation performance \cite{finetuning5, finetuning6}.

\emph{2) Reinforcement learning for preference alignment}

After pre-training and fine-tuning, LLMs gain general generation ability and domain adaptability, yet they may still produce factually incorrect or misaligned outputs. RL therefore serves as a critical post-training stage for aligning model behavior with human preferences and task objectives. Reinforcement learning from human feedback (RLHF) \cite{ouyang2022training} improves response quality, safety, and alignment by incorporating human or learned preference signals into iterative policy optimization, contributing to the success of systems such as ChatGPT and GPT-4 \cite{achiam2023gpt}. As a reward-driven learning paradigm, RL is particularly valuable for optical network O\&M tasks, where decisions often involve trade-offs rather than deterministic answers. For example, feasible solutions in power optimization or fault recovery may satisfy technical constraints but differ in robustness, operational risk, or service impact. Such engineering preferences are difficult to encode through supervised datasets alone but can be naturally incorporated through reward-based learning \cite{liu2026developing}.

\emph{3) Retrieval-augmented generation (RAG)}

Although fine-tuning and RL improve domain adaptation and behavioral alignment, LLMs may still generate hallucinated or incomplete responses when tasks require precise technical knowledge or rigorous physical reasoning. In optical networks, many problems, such as interpreting nonlinear interference (NLI) or reasoning under multi-parameter constraints, depend on specialized knowledge that cannot be reliably encoded into model parameters alone. In this case, RAG addresses this limitation by grounding model outputs in external knowledge sources, improving response accuracy and relevance \cite{RAG1}. Effective RAG depends less on complex architectures than on a well-structured knowledge foundation capable of providing relevant and reliable information when needed \cite{RAG5, RAG6}. For optical networks, this requires a structured knowledge library where physical theories, engineering guidelines, standards, and operational documents are explicitly organized and retrievable, rather than relying solely on the model’s internal knowledge. To ensure efficient use, knowledge should be carefully categorized, structured, and stored in vector form for fast retrieval. When a task is encountered, relevant knowledge is retrieved and incorporated into the LLM prompt, enabling context-aware and domain-grounded reasoning.

\emph{4) Prompt engineering}

By bridging the gap between user intent and model understanding \cite{prompt1}, whether the model is pre-trained, fine-tuned, or reinforced, leveraging prompt engineering can unlock the full potential of LLMs, particularly in domain-specific applications. A well-constructed prompt typically includes four essential elements \cite{prompt2}: instruction, context, input data, and output indicator. The instruction defines the task clearly and steers the model’s reasoning toward user intents. Context supplements the instruction with domain knowledge, external data, or operational constraints, to enhance the model's accuracy. Input data provides task-specific information, enabling the model to handle particular needs effectively. The output indicator specifies the desired format or structure, ensuring usability in downstream workflows. Beyond prompt elements, prompt techniques further enhance the reasoning and analytical capabilities of LLMs \cite{zhao2023survey}. Chain-of-thought (CoT) prompting supports step-by-step reasoning by decomposing complex problems into sequential subtasks \cite{wei2022chain}. This capability is particularly valuable in optical networks, where many operational tasks are inherently multi-step and cannot be solved reliably through a single direct response.

\emph{5) AI Workflow with toolkit}

To enhance the reliability and efficiency of LLMs in task execution, workflow-based systems have emerged as a foundational paradigm. In this framework, complex tasks are decomposed into well-defined sub-steps, where LLMs operate within pre-designed execution pipelines and interact with external modules through standardized interfaces. Unlike early LLM applications that focused primarily on language generation, workflow systems emphasize task-oriented automation, with the LLM acting as an intelligent coordinator within a structured execution process. As illustrated in Fig. \ref{fig-3.C} (e), workflow execution typically involves task decomposition, action planning, execution, and result aggregation, ensuring controllability and reproducibility for stable tasks. Many optical-network operations naturally align with this paradigm. For example, network planning can be decomposed into topology parsing, traffic demand analysis, routing and spectrum assignment (RSA), QoT estimation, and configuration generation. Similarly, operational tasks such as fault localization can be structured into sequential stages including alarm and telemetry collection, anomaly correlation analysis, root-cause inference, fault localization, and recovery recommendation generation. By organizing these steps into modular pipelines, workflow can coordinate domain-specific models and computational modules to accomplish complex engineering tasks more reliably and systematically. Workflow-based execution is particularly suitable when the task objectives, execution steps, required tools, and decision criteria can be specified in advance. In such cases, deterministic workflows provide advantages in repeatability, efficiency, and controllability, and there is no need to introduce autonomous Agent reasoning beyond the predefined procedure.

\emph{6) Harness engineering-enhanced Agent}

While workflow systems offer strong reliability, their predefined structures limit adaptability in dynamic and open-ended environments. To address this limitation, LLM Agents have evolved beyond static tool orchestration toward closed-loop, context-aware, and self-adaptive systems, as illustrated in Fig. \ref{fig-3.C} (f). An Agent is a self-directed entity that leverages LLMs for planning, reasoning, memory management, and tool invocation, enabling iterative action refinement through environmental feedback. Rather than replacing domain-specific tools or deterministic workflows, an Agent provides a flexible reasoning and orchestration layer above them. Given a high-level objective, the Agent can determine which information and capabilities are required, select and invoke appropriate tools or workflows, interpret their intermediate results, and adapt the subsequent execution according to the current context. Compared with workflows, which rely on predefined execution pipelines, Agents dynamically reason and adjust their behaviors at runtime, making them better suited for tasks requiring flexibility and adaptation \cite{amigoni2003anthropic}. Recent advances have further expanded the Agent paradigm through harness engineering, which provides a structured framework for orchestrating and operationalizing Agent behavior.

Harness engineering further strengthens Agent reliability by organizing context management, tool interaction, persistent memory, observation, verification, and recovery mechanisms. These capabilities support closed-loop execution in which an Agent can evaluate intermediate results, refine its actions, and coordinate with other Agents through multi-Agent or Agent-to-Agent (A2A) collaboration \cite{hong2023metagpt, wu2024autogen}. Unlike static workflows, harness-enhanced Agents can dynamically determine task decomposition and execution strategies at runtime, making them well suited for complex optical network O\&M scenarios.

\section{LLM-Centric Full-Lifecycle Management for Agentic Optical Network}

\begin{figure*}[t]
\centering
\includegraphics[width=14cm]{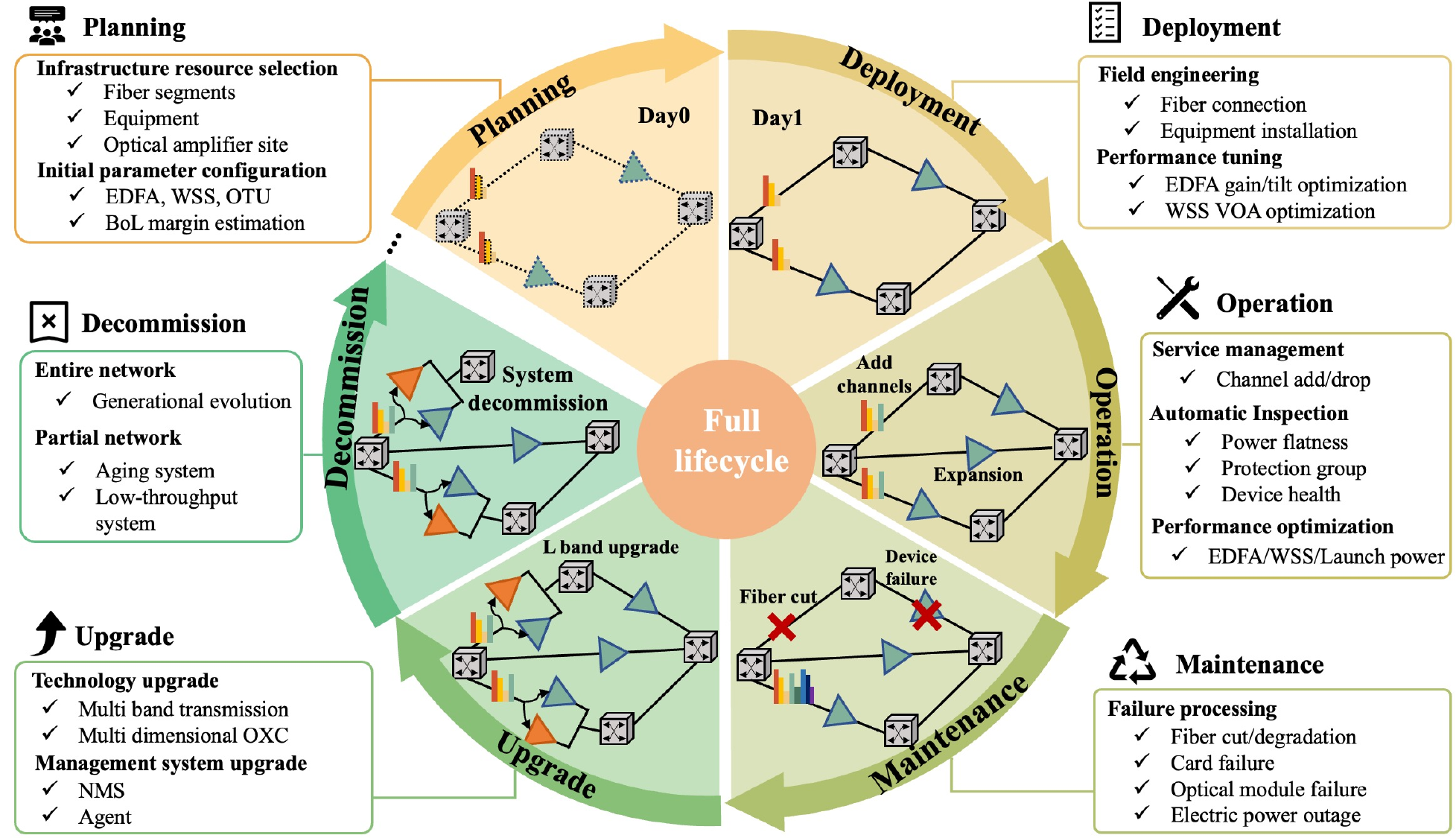}
\caption{Diagrammatic representation of the key tasks corresponding to each phase of the optical network lifecycle, covering planning, deployment, operation, maintenance, upgrade, and decommission.}
\label{fig-4.A}
\end{figure*}

The full LCM of optical networks encompasses multiple phases, including planning, deployment, operation, maintenance, upgrade, and decommissioning, typically spanning several decades, as illustrated in Fig. \ref{fig-4.A}. As the foundational layer of modern communication infrastructure, enhancing the level of autonomy in optical networks is critical for improving the reliability of cross-domain traffic delivery while reducing the operational complexity associated with large-scale network management. In this context, LLMs, with their strong capabilities in intent understanding and logical reasoning, have the potential to support a wide range of O\&M tasks. More importantly, LLMs can be envisioned as the cognitive core of AONs, continuously evolving alongside the network and permeating all lifecycle phases. Consequently, future AONs are expected to emerge as unified and intelligent systems, jointly constructed by physical infrastructure, DT, control planes, and multi-Agent systems.

To systematically enable LLM-centric LCM in AONs, it is essential to first establish a comprehensive understanding of the tasks and workflows across different lifecycle phases. Accordingly, this section summarizes the major responsibilities of each phase, followed by the proposed hierarchical multi-Agent framework for full-lifecycle automation.

\subsection{The full lifecycle of optical networks}

\noindent \textit{1)	Planning}

The planning phase represents the initial stage of the optical network lifecycle, which involves defining network requirements, forecasting future capacity needs, and creating a blueprint for the physical and logical architecture \cite{ramaswami2001optical}. This phase includes analyzing bandwidth demands, coverage targets, and service types, selecting appropriate technologies, designing fiber routes and redundancy schemes, and planning node placement across core, aggregation, and access layers. Equipment selection must balance performance, cost, and supply chain robustness, while operators often adopt multi-vendor strategies to mitigate vendor lock-in and enhance network flexibility. Based on candidate fiber routes and equipment options, optical amplifiers (OAs) placement and optical transponder units (OTUs) allocation can be jointly optimized with fiber and equipment selection. The corresponding decisions may include the locations and quantities of OAs and the allocation, types, and capacities of OTUs, subject to fiber-loss, QoT, cost, reliability, and deployment constraints.

Building upon the selected resources, a preliminary network topology is constructed, enabling subsequent performance evaluation based on both fiber characteristics and device parameters. This evaluation provides the foundation for downstream configuration tasks, including wavelength planning driven by initial bandwidth demands. The outputs of the planning phase therefore comprise an optimized resource allocation and deployment scheme, initial configuration parameters for key optical devices, such as OAs, WSSs, and OTUs, as well as a unified IP addressing plan. This phase, often referred to as the Day 0 stage of the optical network lifecycle, establishes the baseline for all subsequent deployment and operational processes.

\noindent \textit{2)	Deployment}

Once the planning phase is completed, the optical network transitions into the deployment phase, during which the designed infrastructure is physically instantiated and initialized for operation. This phase encompasses fiber connectivity establishment, equipment installation and configuration, as well as initial system performance optimization. Building upon the established physical connectivity and baseline configurations, the NMS subsequently assumes control, enabling real-time monitoring and data collection to support the construction of DT \cite{wang2022review, zhuge2023building}. Based on the acquired network state information, optimization algorithms can be applied to validate and refine system performance, ensuring that service requirements are satisfied. Meanwhile, key deployment-stage data, which including fiber parameters, configuration files, begin-of-life (BoL) performance margins, and physical device location information, are systematically recorded to provide a reliable reference for subsequent operation stages. Upon completion of these processes, the network becomes service-ready, marking the transition to the Day 1 stage of the optical network lifecycle.

\noindent \textit{3)	Operation}

Following deployment, the optical network transitions into the operation phase, during which it continuously carries and manages live IP traffic. This phase is inherently dynamic and long-term \cite{musumeci2018overview}, involving three tightly coupled functional processes: adaptive optical channel management, continuous network health assessment, and real-time performance optimization. As service demands evolve over time, optical channels must be dynamically provisioned, adjusted, or released to accommodate traffic fluctuations. In parallel, network health is continuously evaluated through the analysis of performance metrics, sensor readings, and device status information. Based on the observed network state, performance optimization mechanisms are further triggered to maintain service quality \cite{pointurier2016design}.

Compared to the planning and deployment phases, the operation phase exhibits significantly higher complexity and persists over a longer period while continuously interacting with live traffic and facing the stringent requirement for ultra-reliable execution, where even minor disruptions may lead to substantial service impact. This necessitates decision-making processes to be not only accurate but also risk-aware and globally coordinated. In this context, autonomous operation emerges as a critical enabler, capable of reducing human-induced errors while rapidly adapting to unexpected network conditions through holistic, system-level awareness.

\noindent \textit{4)	Maintenance}

Maintenance is performed concurrently with the operation phase and represents the most enduring stage of the optical network lifecycle, playing a critical role in sustaining long-term network reliability and availability. This phase focuses on the timely detection, localization, and mitigation of network faults arising from diverse sources \cite{wang2022review, musumeci2018overview}. To address such failures, maintenance activities rely on a combination of network monitoring and coordinated field operations. In NMS, alarms are generated to facilitate fault analysis and location, after which maintenance personnel collaborate with field technicians to perform recovery actions, including fiber repair and module replacement. The efficiency of fault detection and resolution directly impacts the overall availability and continuity of services in the network. However, in large-scale optical networks, traditional manual maintenance mode faces significant challenges in terms of scalability, response time, and knowledge consistency, particularly in the presence of staff turnover and increasingly complex system configurations. In this context, autonomous maintenance is anticipated to be a key enabler to accelerate fault recovery, reduce operational overhead, and alleviate reliance on human expertise.

\noindent \textit{5)	Upgrade}

The upgrade phase of optical networks encompasses both technological evolution and network scale expansion, enabling the system to adapt to the long-term growth in service demand and advances in transmission and control capabilities. Such upgrades may involve capacity expansion, equipment replacement, spectrum reconfiguration, and updates to control and management functions as network requirements evolve. Despite occupying a relatively small portion of the entire lifecycle, the upgrade phase is inherently high-risk, as any misconfiguration or maloperation may lead to performance degradation and even service disruption in live networks. Consequently, upgrade operations require meticulous planning and evaluation, typically involving extensive simulations and performance assessments on the DT platform, and in many cases, physical testing within controlled laboratory environments. In this context, the integration of high-level autonomy with DT technologies is expected to significantly improve both the efficiency and reliability of upgrade processes by enabling predictive analysis, risk-aware decision-making, and pre-deployment validation.

\begin{figure*}[ht]
\centering
\includegraphics[width=15cm]{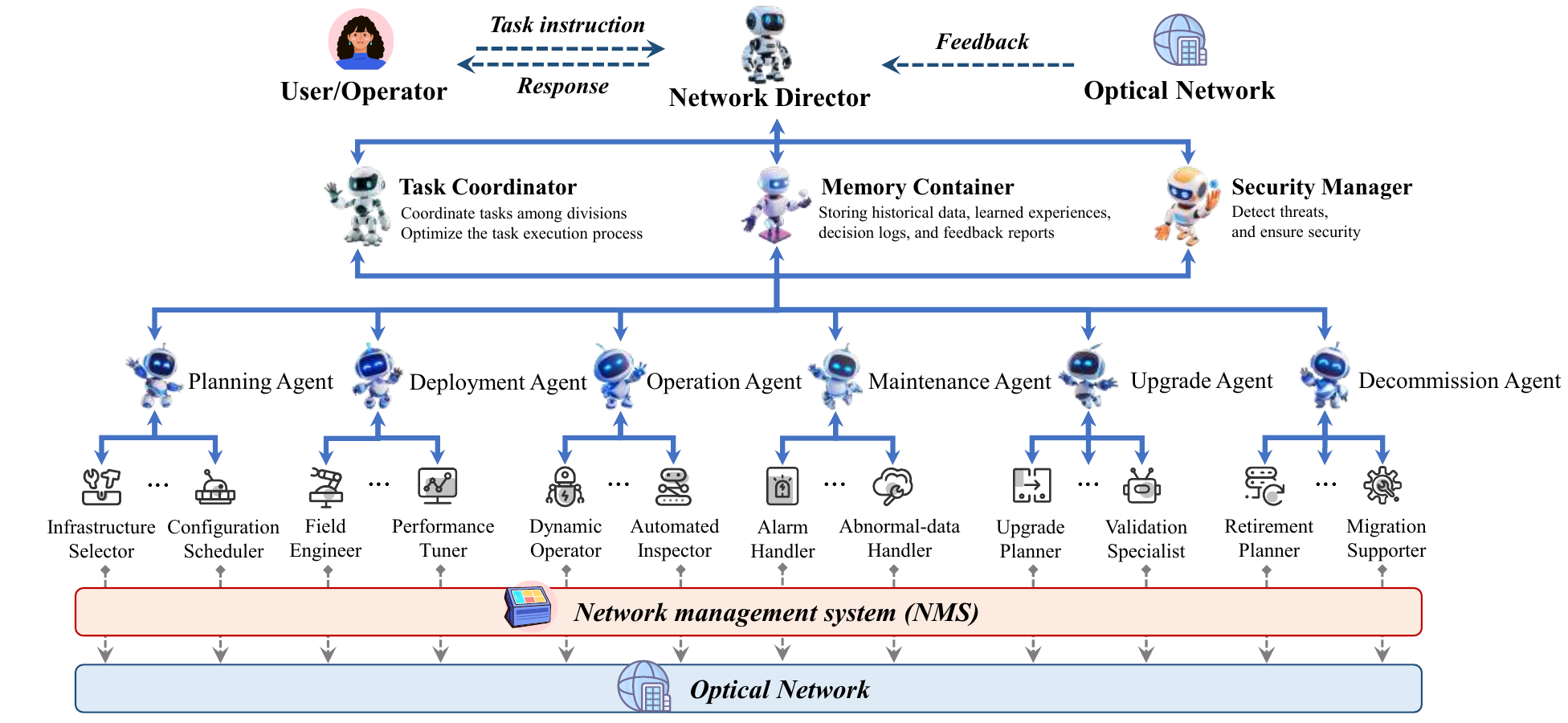}
\caption{Hierarchical multi-Agent framework for optical network full-lifecycle management. The architecture consists of four layers of intelligent Agents. At the top layer, the Network Director Agent interacts with the user/operator, handling both intent input and feedback aggregation from lower layers. The second layer comprises three administrative Agents, each responsible for a major domain of network lifecycle management. The third layer includes primary Agents that oversee the automation of specific lifecycle phases. At the bottom layer, sub-Agents are assigned to execute concrete tasks.}
\label{fig-5.A framework}
\end{figure*}

\noindent \textit{6)    Decommission}

The final phase of the optical network lifecycle is decommissioning, which marks the systematic retirement of network infrastructure as it becomes obsolete or no longer economically viable. This phase is typically triggered by several factors, including large-scale technology transitions, progressive system aging, and declining service demand that renders existing systems inefficient \cite{takita2017towards}. Decommissioning is inherently complex, involving large-scale service migration, resource reclamation, and coordinated asset management across multiple network layers. Among these processes, traffic cutover operations are particularly critical, as they must be executed with minimal disruption to ongoing services. Consequently, ensuring a safe and seamless transition requires precise planning, real-time monitoring, and tightly coordinated execution. In this context, autonomous network capabilities play a pivotal role by enabling automated service migration, reducing operational risks, and maintaining service continuity throughout the decommissioning process.

Across these phases, the network evolves from design and physical realization to long-term operation, evolution, and eventual retirement. The diversity of objectives, constraints, and operational conditions across the lifecycle motivates a phase-oriented organization of Agent responsibilities, as discussed in the following subsection.

\subsection{Hierarchical multi-Agent framework}

To effectively facilitate the autonomous LCM of optical networks, we propose a hierarchical multi-Agent framework comprising the four levels: 1) a Network Director overseeing global operations, 2) three Administrative Agents providing cross-cutting capabilities for task coordination, memory management, and security support, including Task Coordinator, Memory Container, and Security Manager, 3) six Division Agents responsible for different phases of the network lifecycle, including planning, deployment, operation, maintenance, upgrade, and decommission phases, and 4) multiple AI Experts within each division executing specific functions. Each Agent is designed to autonomously execute tasks within its domain while collaborating across layers, ensuring seamless and intelligent optical network management, as depicted in Fig. \ref{fig-5.A framework}.

\begin{table*}[t]
\centering
\caption{Representative knowledge, network data, and tools for Agents across different lifecycle phases of AONs.}
\label{knowledge_tools}
\renewcommand{\arraystretch}{1.5}
\setlength{\tabcolsep}{4pt}
\footnotesize
\begin{tabular}{p{2cm} p{7cm} p{7cm}}
\hline
\textbf{Lifecycle phase} &
\textbf{Representative knowledge and data} &
\textbf{Representative tools and interfaces} \\
\hline

\textbf{Planning} &
Network topology; traffic demands and forecasts; equipment specifications; deployment constraints; equipment databases; &
RSA tool; QoT estimation; resource planning; DT platform; \\

\textbf{Deployment} &
Equipment inventory; device capabilities; configuration templates; installation records;  &
NMS API; device configuration interfaces; commissioning and diagnostic tools; \\

\textbf{Operation} &
service statistics; topology and network state; QoT measurements; alarm information; action logs; &
Real-time telemetry APIs; QoT estimation; RSA tool; DT platform; NMS API; performance-analysis scripts; \\

\textbf{Maintenance} &
Alarm rules; historical cases; maintenance records; OTDR traces; equipment state information; troubleshooting procedures; &
Alarm analysis tools; fault-diagnosis tools; OTDR APIs; DT platform; NMS API; recovery and reconfiguration tools; \\

\textbf{Upgrade} &
Capacity requirements; traffic growth trends; equipment databases; spectrum utilization; upgrade constraints; 
&
Capacity optimization tool; RSA tool; QoT estimation; DT platform; resource planning; configuration and validation tools; \\

\textbf{Decommission} &
Service inventory; SLA requirements; equipment and asset records; retirement procedures; &
QoT and performance verification; service migration tools; NMS API; asset-management systems; configuration tools \\

\hline
\end{tabular}
\end{table*}

At the highest level of the framework, the Network Director serves as the central decision-making entity, orchestrating the entire multi-Agent system. It acts as the primary interface between human operators and the AI Agents, maintaining a global view of network objectives and operational states. It is initially customized with basic knowledge of optical network operation and is responsible for controlling the overall task execution process of network automation. Rather than directly executing all domain-specific tasks, the Network Director interprets high-level intents, allows the system to maintain a unified network-wide objective while delegating specialized tasks to dedicated Agents.

At the secondary level, three Administrative Agents support the Network Director: Task Coordinator, Memory Container, and Security Manager, each responsible for a critical aspect of autonomous network governance. These Agents provide cross-cutting capabilities that are shared by multiple lifecycle divisions and therefore are not tied to any particular network phase. The Task Coordinator is responsible for the distribution of tasks across divisions and optimizing the task execution process. The Memory Container functions as a knowledge repository, storing historical data, learned experiences, decision logs, and feedback reports, thereby preserving contextual information across different lifecycle phases and supporting informed decision-making. The Security Manager ensures network reliability and safety through proactively monitoring anomalies and faults, detecting potential threats, and pre-validating configuration strategies to protect the system from risks. For actions that may affect the physical network, the Security Manager can validate strategies from a global perspective before execution and triggering human confirmation for high-risk operations. Separating these shared functions from individual Division Agents avoids duplicated coordination, memory, and security mechanisms and enables consistent policies and contextual information to be maintained throughout the network lifecycle.

At the division level, six Division Agents designed for different phases of the optical network: Planning, Deployment, Operation, Maintenance, Upgrade, and Decommission, are responsible for managing the full lifecycle of the optical network. This phase-oriented organization assigns a clear ownership of tasks according to their lifecycle context, since different phases have distinct objectives, constraints, information requirements, and decision processes. Each Division Agent acts as a middle manager that interprets task targets from administrative layers, formulates phase-specific strategies, and orchestrates task execution within its domain. It also validates the outputs of its sub-Agents against task objectives and constraints, forming a local closed loop in which unsatisfactory results are refined or re-executed before being passed to the upper level. These Agents ensure that every lifecycle phase, from initial planning to network decommission, is intelligently coordinated and seamlessly integrated into the overall management process.

At the expert level, AI Experts operate under the supervision of their respective Division Agents to execute specific sub-tasks efficiently. The Planning Agent focuses on network design, resource allocation, and feasibility analysis. It can direct the Infrastructure Selector to choose appropriate components and instruct the Configuration Scheduler to implement initial settings. The Deployment Agent ensures seamless infrastructure rollout, commanding the Field Engineer for physical installation and the Performance Tuner for system verification and stability testing. The Operation Agent maintains real-time network functionality, with the Automated Inspector executing anomaly detection and the Dynamic Operator managing traffic and resource allocation according to the actual requirements. The Maintenance Agent ensures reliability by directing the Alarm Handler to repair problems, and the Abnormal-data Handler to prevent failures. The Upgrade Agent enhances performance, instructing the Upgrade Planner to implement improvements and the Validation Specialist to verify network stability. The Decommission Agent manages network retirement, overseeing the Retirement Planner for equipment removal and the Service Switcher for seamless data and service transitions. Experts shown here are representative rather than exhaustive; additional task-specific Experts, such as Device Manager and Physical Decommission Agent, can be instantiated according to specific network requirements and may be added as the network evolves. This division-to-expert structure separates lifecycle-level coordination from task-level execution, allowing each Expert to specialize in a specific function and interact with the corresponding models, knowledge bases, and network tools.

\begin{figure*}[t]
\centering
\includegraphics[width=16.5cm]{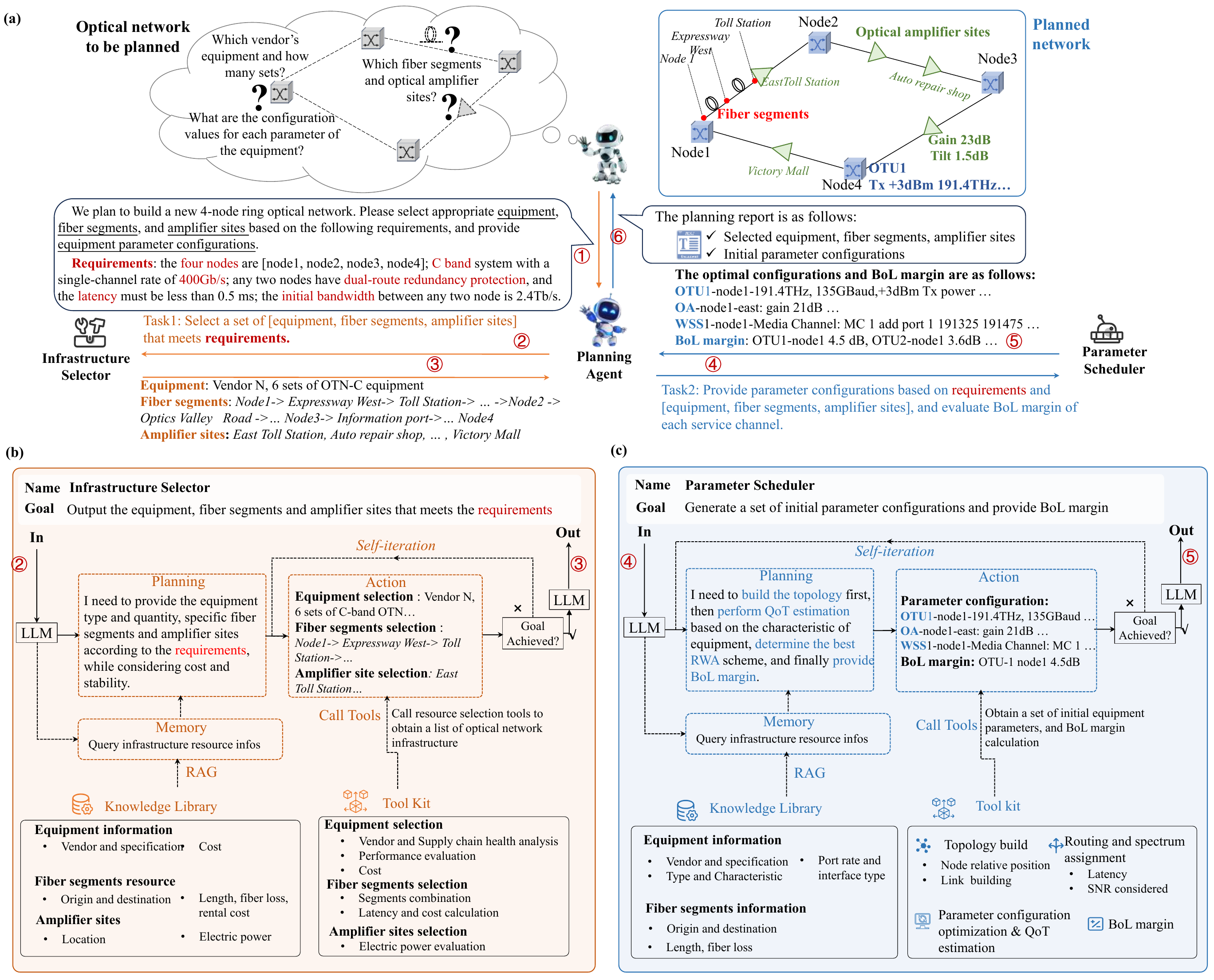}
\caption{Schematic of multi-Agent collaboration in the planning phase of the optical network, (a) coordination among Planning Agent, Infrastructure Selector, and Parameter Selector, (b)-(c) functional workflow of the Infrastructure Selector and the Parameter Scheduler: based on the input, the LLM plans the task, retrieves necessary information via RAG from the knowledge base, and invokes tools to carry out a sequence of operations. Once the task is completed and the objective is met, it returns the results to the higher-level Agent.}
\label{fig-5.1 planning}
\end{figure*}

As a whole, all Agents are capable of inter-Agent communication via A2A protocols \cite{google2025a2a_github} and interact synergistically to ensure seamless coordination across the entire network lifecycle. The Planning, Deployment, and Decommission Agents are activated when their corresponding lifecycle phases are initiated: the Planning Agent defines network architecture and resource allocation, the Deployment Agent manages infrastructure rollout and initial setup, and the Decommission Agent oversees retirement and migration. In contrast, the Operation, Maintenance, and Upgrade Agents remain continuously active throughout the network's operational phase, ensuring real-time monitoring, maintenance, and performance optimization. These three Agents form the core of network management, working closely with the Task Coordinator, Memory Container, and Security Manager to maintain efficiency, adaptability, and security. For example, when the Operation Agent detects abnormal performance metrics, it first attempts self-correction before escalating the issue to the Task Coordinator, which then assigns the task to the Maintenance Agent for immediate troubleshooting or to the Upgrade Agent for hardware adjustments and system enhancements.

Although all these multi-level Agents are powered by LLMs as the core technology, we can flexibly select models of different scales based on task complexity and capability requirements, enabling hierarchical deployment across the cloud, edge, and device. Core control and complex task processing can be handled by large models in the cloud, while latency-sensitive and resource-constrained tasks are efficiently executed by relatively small models deployed at the edge or devices, optimizing computational resource allocation while ensuring high-level intelligence. In practical deployment, each lifecycle Agent is grounded in a combination of domain knowledge, network-state information, and executable tools. Knowledge and data provide the contextual information required for Agent reasoning, while domain-specific tools provide validated computational, simulation, monitoring, and control capabilities. The Agent coordinates these heterogeneous resources according to the task and the current network state. Representative knowledge sources, network data, and tools for the six lifecycle phases are summarized in Table \ref{knowledge_tools}.

\section{Autonomous Lifecycle Management of Agentic Optical Networks}

The AONs extend beyond single task automation toward coordinated and lifecycle intelligence. In this section, we present the architecture of LLM Agent-driven autonomous LCM of optical networks, encompassing the detailed workflow of the six phases from planning to decommission. Instead of focusing on individual task automation, we emphasize how multiple specialized Agents can be organized and orchestrated to support all stages in a coherent and scalable manner. By introducing how Agent capabilities can be integrated into optical networks, we aim to provide a structured foundation that can inform initial system design and deployment for future AONs.

\subsection{Planning phase}

In the initial planning phase of the optical network lifecycle, the primary task is to perform demand analysis and topology design based on factors such as the operator’s service development plan, capital investment strategy, and network reliability requirements. This process determines the fundamental attributes of the new network, including node geographic locations, network capacity, redundancy protection schemes, latency requirements between nodes, and initial bandwidth provisioning. As this process constitutes the top-level design of the network, it involves diverse and dynamically changing information elements, and it is necessary to combine information from a knowledge base and utilize various tools for Agents. To address these challenges, based on the above hierarchical multi-Agent framework, it is designed to deploy three Agents for the planning phase. First, the Infrastructure Selector Agent is responsible for selecting infrastructure resources. Second, the Parameter Scheduler Agent generates configuration parameters. Third, the Planning Agent serves as the orchestrator, coordinating task reception, information integration, and downward command delegation, functioning as a communication bridge within this multi-Agent system.

These roles do not imply that every step requires LLM reasoning. Deterministic database queries, rule-based filtering, QoT estimation, and numerical optimization remain preferable for well-defined inputs and constraints because they provide reliable and reproducible outputs. The added value of LLM-based Agents lies at the orchestration level, particularly in interpreting high-level or incomplete operator intents, translating contextual preferences into selection criteria and optimization objectives, reasoning over trade-offs among performance, cost, supply-chain robustness, and reliability, and dynamically selecting and sequencing the required tools. Accordingly, the Infrastructure Selector uses LLM reasoning to synthesize heterogeneous information and establish resource-selection priorities, whereas the Parameter Scheduler invokes deterministic tools to generate reproducible configurations and performance estimates. For stable tasks with fixed requirements and predefined workflows, a purely deterministic implementation may be sufficient. Agent-based orchestration becomes valuable when requirements are incomplete or evolving, multiple tools must be coordinated, or intermediate results require the workflow to be adapted.

\begin{figure*}[t]
\centering
\includegraphics[width=16.5cm]{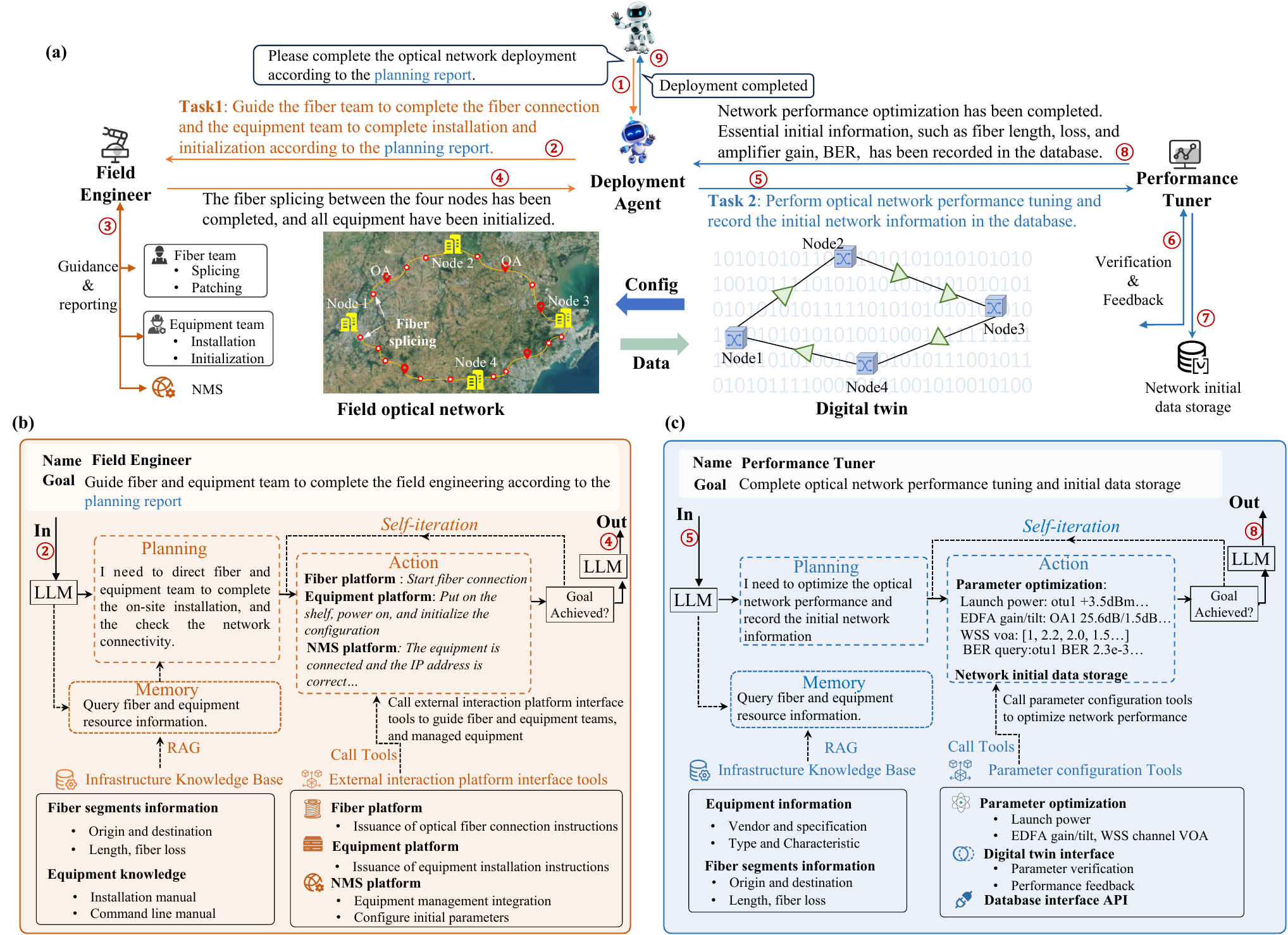}
\caption{Schematic of multi-Agent collaboration in the deployment phase of the optical network, (a) coordination among Deployment Agent, Field Engineer, and Performance Tuner, (b) functional workflow of the Field Engineer, (c) functional workflow of the Performance Tuner.}
\label{fig-5.2 deployment}
\end{figure*}

In this phase, Network Director notifies the Task Coordinator to process the user-defined task and assign the optical network planning task to the Planning Agent. Taking a simple four-node ring topology as an example, the collaboration process among these Agents can be illustrated as depicted in Fig. \ref{fig-5.1 planning} (a). First, the Planning Agent receives key information about the network, including the names of four nodes, the per-wavelength rate of a C-band system, and the requirement for dual-path redundancy protection. Based on these requirements, the Planning Agent delegates the task of resource selection to the Infrastructure Selector. The workflow of the Infrastructure Selector Agent is depicted in Fig. \ref{fig-5.1 planning}(b). After understanding the Planning Agent’s intent, the Infrastructure Selector retrieves information on fiber segments related to the specified node locations and corresponding amplifier sites using RAG from the infrastructure knowledge base. Meanwhile, it retrieves equipment information that meets user requirements, including vendor names, technical specifications, and pricing. Next, the Infrastructure Selector enters the Action phase and invokes resource-selection and deployment-optimization tools to select equipment and fiber segments and, where multiple candidate solutions are available, to determine the locations and quantities of OAs and the allocation of OTUs. Next, the Infrastructure Selector enters the Action phase, invoking resource selection tools to choose from the three categories of resources: equipment, fiber segments, and amplifier sites. Equipment selection must consider multiple criteria such as supply chain robustness, cost, and technical performance, with specific priorities depending on the operator’s strategic preferences. Once a feasible set of resources is selected and deemed to meet user requirements, the Infrastructure Selector sends the result to the Planning Agent. If the selected resources fail to meet expectations, the Infrastructure Selector must iterate on the selection process.

Following this, the Planning Agent integrates user requirements with the selected infrastructure resources and instructs the Parameter Scheduler to generate initial configuration parameters and estimate BoL margin for service channels. Upon receiving instructions from the Planning Agent, Parameter Scheduler initiates its workflow by retrieving detailed resource information via RAG from the knowledge base. It then proceeds to invoke a series of tools: first, a topology construction tool used to simulate the four-node ring network, followed by a QoT estimation tool and device configuration optimization algorithm. These tools determine parameters such as OA gain/tilt and WSS channel attenuation settings. The Parameter Scheduler then completes RSA according to latency and SNR requirements. Finally, the BoL margin of the service channels is obtained. If the margin values meet or exceed the minimum threshold, the Parameter Scheduler submits the generated configuration back to the Planning Agent. After that, Planning Agent consolidates the final planning results into a planning report, which includes a detailed network topology and configuration summary. This report is returned to the Task Coordinator, which guides the subsequent deployment phase.

This example illustrates how three Agents can collaborate to efficiently complete infrastructure selection, parameter pre-generation, and performance estimation for the planning phase. It is important to note that all Agents are based on LLM and rely heavily on well-designed prompts and harness engineering to effectively execute planning, action reasoning, RAG queries, tool invocation, and result evaluation. Additionally, to ensure optimal infrastructure selection by the Infrastructure Selector Agent, the completeness and standardization of information in the infrastructure knowledge base is critical. Similarly, for the Parameter Scheduler Agent, beyond knowledge base quality, the accuracy and reliability of tools in the toolset library are equally essential. In this framework, LLM-based Agents act as orchestrators rather than directly replacing deterministic numerical optimizers. They translate planning requirements into objectives and constraints, retrieve candidate resources and uncertainty information, invoke appropriate deterministic or robust optimization tools, and verify and iterate the returned solutions. Robust or rolling-horizon optimization can also be invoked when traffic demand, component aging, failure risks, or resource availability evolve over time. Therefore, OA and OTU deployment optimization can be implemented as a specialized computational module while remaining embedded as a tool-enabled subtask of the planning phase. The resulting deployment plan is then physically implemented in the subsequent deployment phase.

\subsection{Deployment phase}

In the deployment phase of the optical network lifecycle, the primary objective is to implement the network topology and equipment configurations generated during the planning phase and to perform initial performance tuning. This phase is characterized by frequent interactions between the network operation center (NOC) and field engineers, as well as communications between the NMS and physical devices. To efficiently manage this complicated task, three types of AI Agents are assigned to this phase: a top-level Deployment Agent, a Field Engineer Agent responsible for field coordination, and a Performance Tuner Agent responsible for network performance optimization.

\begin{figure*}[t]
\centering
\includegraphics[width=15cm]{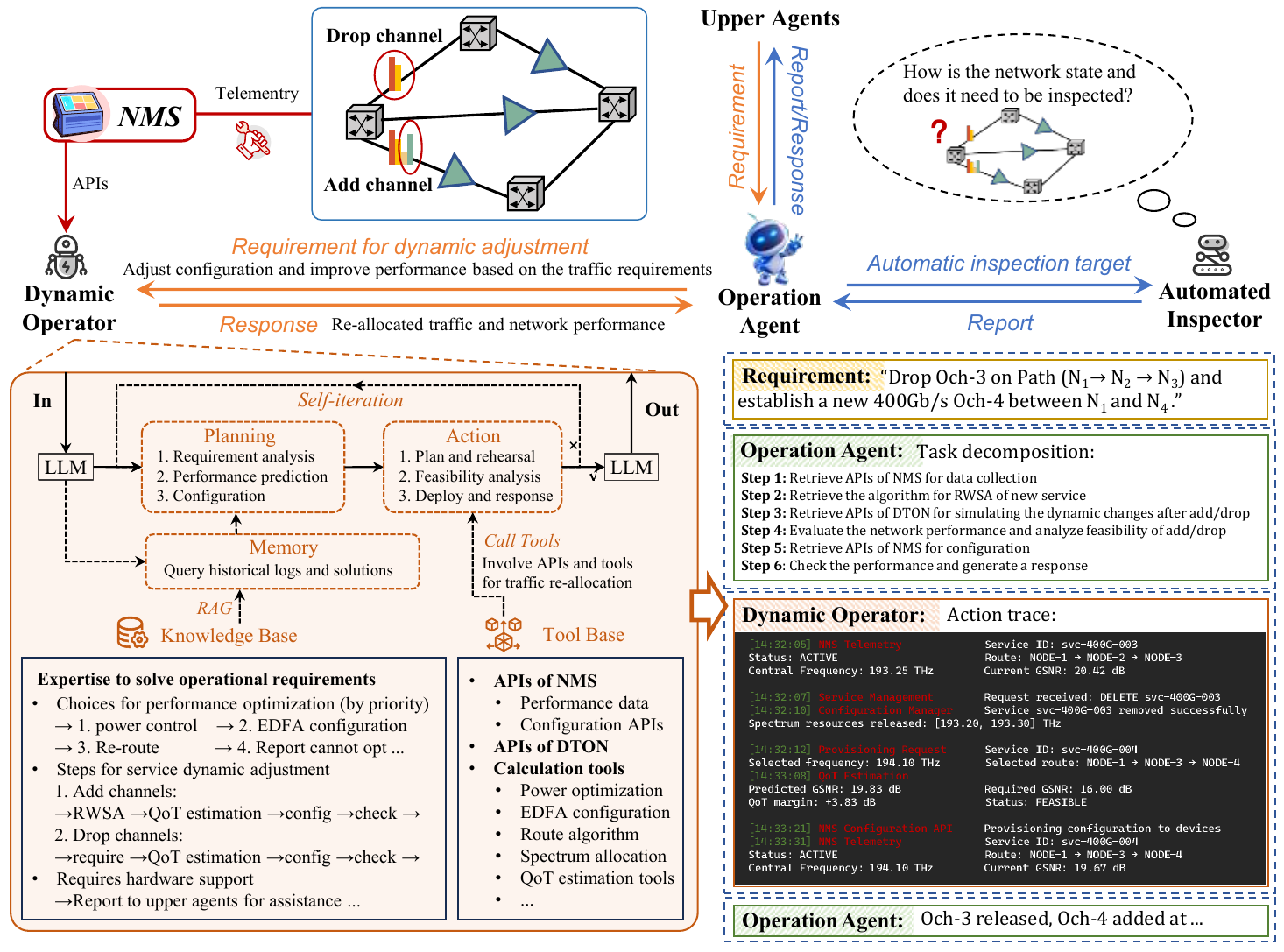}
\caption{An illustration of multi-Agent collaboration during the operation phase of AONs, along with the functional workflow of the Dynamic Operator and a detailed description of its behavior in a channel adding scenario.}
\label{fig-5.3 operation1}
\end{figure*}

The Deployment Agent receives deployment tasks from the Task Coordinator, along with the planning report produced by the Planning Agent. Following the typical workflow of the deployment phase, the Deployment Agent first instructs the Field Engineer Agent to carry out on-site tasks such as fiber connection and equipment installation. Here, the Field Engineer Agent can be embodied and equipped by human with mobile terminals, or potentially through robotic systems for assisted or automated operations. Then Deployment Agent assigns the Performance Tuner to execute optical network performance tuning and store initial network data. The workflow of Field Engineer is illustrated in Fig. \ref{fig-5.2 deployment} (b). Upon receiving instructions, the Field Engineer can access relevant fiber and equipment information through RAG, and interacts with the fiber and equipment platform to coordinate with the fiber and equipment teams, respectively, for field implementation. Since the Field Engineer can interface directly with human field engineers, the platforms may support multimodal interactions including text, voice, image, and video, similar to features provided by modern instant messaging tools. During the fiber connection process, Field Engineer guides the fiber team to connect fiber segments based on the planning report. Some segments are connected via splicing, while others may require patch cords. It is essential that Field Engineer instructs the fiber team to use equipment such as optical time domain reflectometer (OTDR) to ensure fiber attenuation and splice loss meet engineering standards. Additionally, using installation manuals retrieved from the equipment knowledge base, Field Engineer guides the equipment team in installing devices at optical amplifier sites and ROADM nodes. This includes completing fiber patching and initializing the equipment to enable proper inter-device communication. Field Engineer may then invoke the NMS to incorporate the equipment into the management system and deploy the OTU and optical equipment configurations generated during planning. Once these steps are completed, the optical channel (OCH) can be established, and the network reaches a near-optimal operational state. At this point, telemetry data, such as network status and transmission performance, can be collected to construct an DT of the deployed physical network, as shown in Fig. \ref{fig-5.2 deployment} (a).

Since actual fiber parameters may slightly deviate from those assumed during planning, fine-tuning of device configurations is necessary to achieve optimal power flatness and SNR. These adjustments, performed by the Performance Tuner, can be verified within the DT before being issued to the physical network. The optimization parameters mainly includes launch power profile, EDFA gain/tilt, and WSS channel attenuation settings. Once performance metrics converge to their optimal values, the optimized configuration is deployed to the network devices. Finally, the fiber parameters, equipment configurations, and transmission performance data are stored in the network initial database, which serves as the starting point of the lifecycle-wide database for subsequent operation and maintenance phases.

\begin{figure*}[t]
\centering
\includegraphics[width=16.5cm]{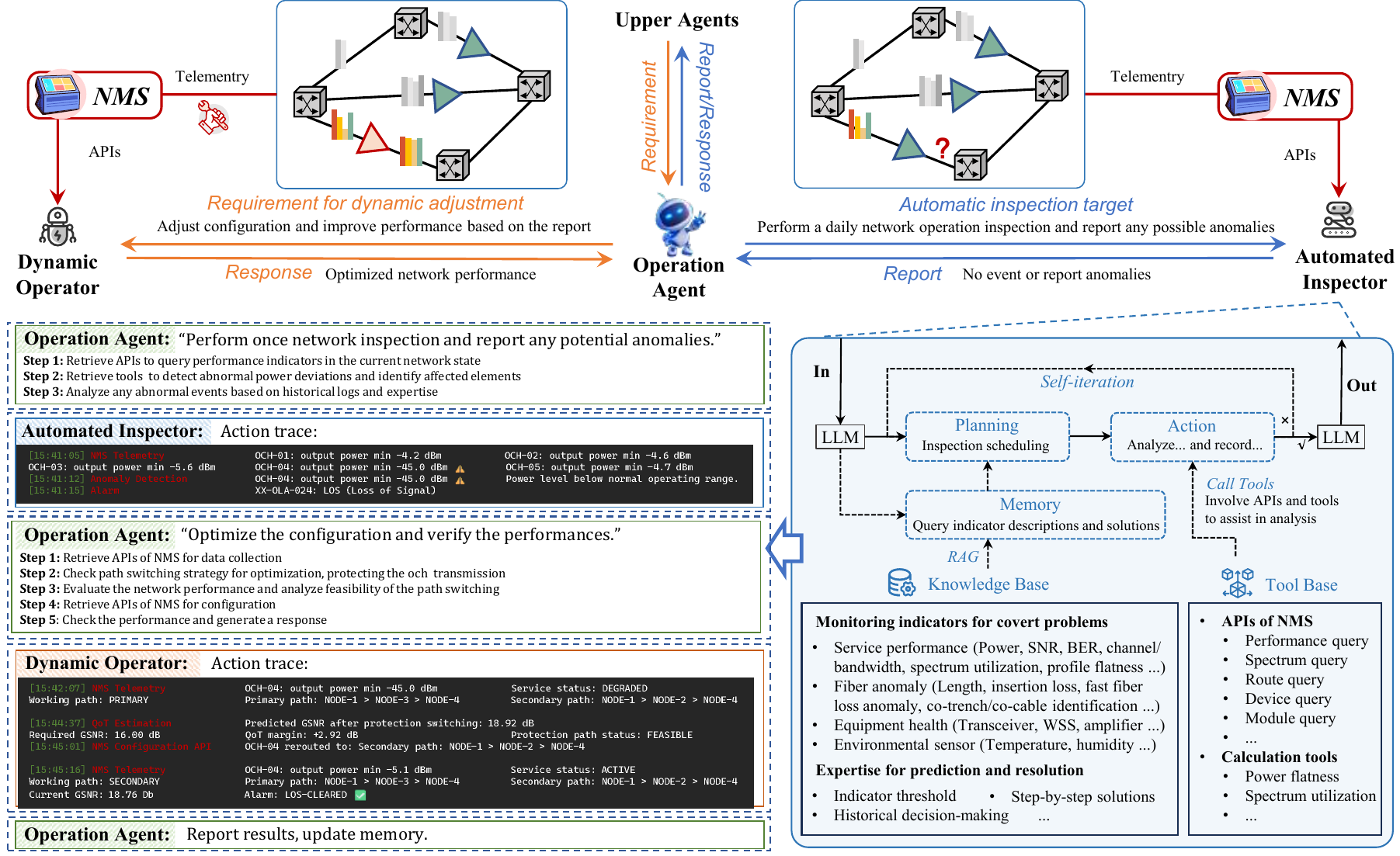}
\caption{Schematic of functional workflow of the Automated Inspector in operation phase, along with a detailed description of its behavior in a network anomaly inspection scenario.}
\label{fig-5.3 operation2}
\end{figure*}

This phase further demonstrates the critical importance of high-quality knowledge and tool libraries in enabling Agent efficiency and accuracy. For example, the Performance Tuner relies heavily on the accuracy of the DT and the effectiveness of optimization algorithms. Therefore, careful design and continuous maintenance of both the knowledge base and toolset are essential throughout the entire network lifecycle.

\subsection{Operation phase}

In the operation phase, the primary focus shifts to ensuring the network’s ongoing stability, efficiency, and reliability, as well as adapting to real-time changes in traffic and condition. This phase is essential for maintaining high-quality service throughout the network's lifecycle. The Operation Agent plays a central role in managing this phase, orchestrating network tasks, monitoring optical performance, and taking corrective actions when necessary. Unlike the Planning Agent and Deployment Agent, which operate only at the initial phases of the network’s lifecycle, the Operation division remains continuously active, overseeing the daily functioning of the network and ensuring that it runs smoothly throughout its operational lifetime.

To efficiently manage the various tasks required during this phase, the Operation Agent works in close collaboration with two specialized sub-Agents: Dynamic Operator and Automated Inspector. Dynamic Operator is responsible for the continuous adjustment of network parameters to accommodate fluctuations in network traffic, bandwidth demands, and other dynamic variables. This Agent makes real-time decisions to balance network traffic and improve performance, ensuring that each part of the network is operating at peak efficiency. It plays a crucial role in continuously optimizing network performance, ensuring service quality, and dynamically adjusting network configurations based on evolving demands. 

One of Dynamic Operator’s primary functions is network performance enhancement to guarantee high-quality service provisioning. It continuously evaluates network performance by analyzing key indicators such as latency, power, SNR, and BER. Based on changes in traffic demand or new service requirements, it can autonomously adjust network configurations and expand capacity as needed. Through NMS, Dynamic Operator primarily executes software-level adjustments, including launch power optimization, device parameter configuration, and routing adjustments. As illustrated in Fig. \ref{fig-5.3 operation1}, a representative dynamic add/drop task demonstrates the execution process from an upper-level requirement to concrete network actions. The Operation Agent first decomposes the requirement into API retrieval, RSA, DT-based performance evaluation, and configuration verification steps. The Dynamic Operator then invokes the corresponding NMS and computational tools and produces an executable action trace. In the demonstrated case, an existing 400 Gb/s service is released and a new service is provisioned on an alternative path and frequency, followed by GSNR evaluation and NMS configuration. The returned network status confirms that the new service is active and that the required transmission performance is satisfied.

\begin{figure*}[t] 
\centering
\includegraphics[width=17cm]{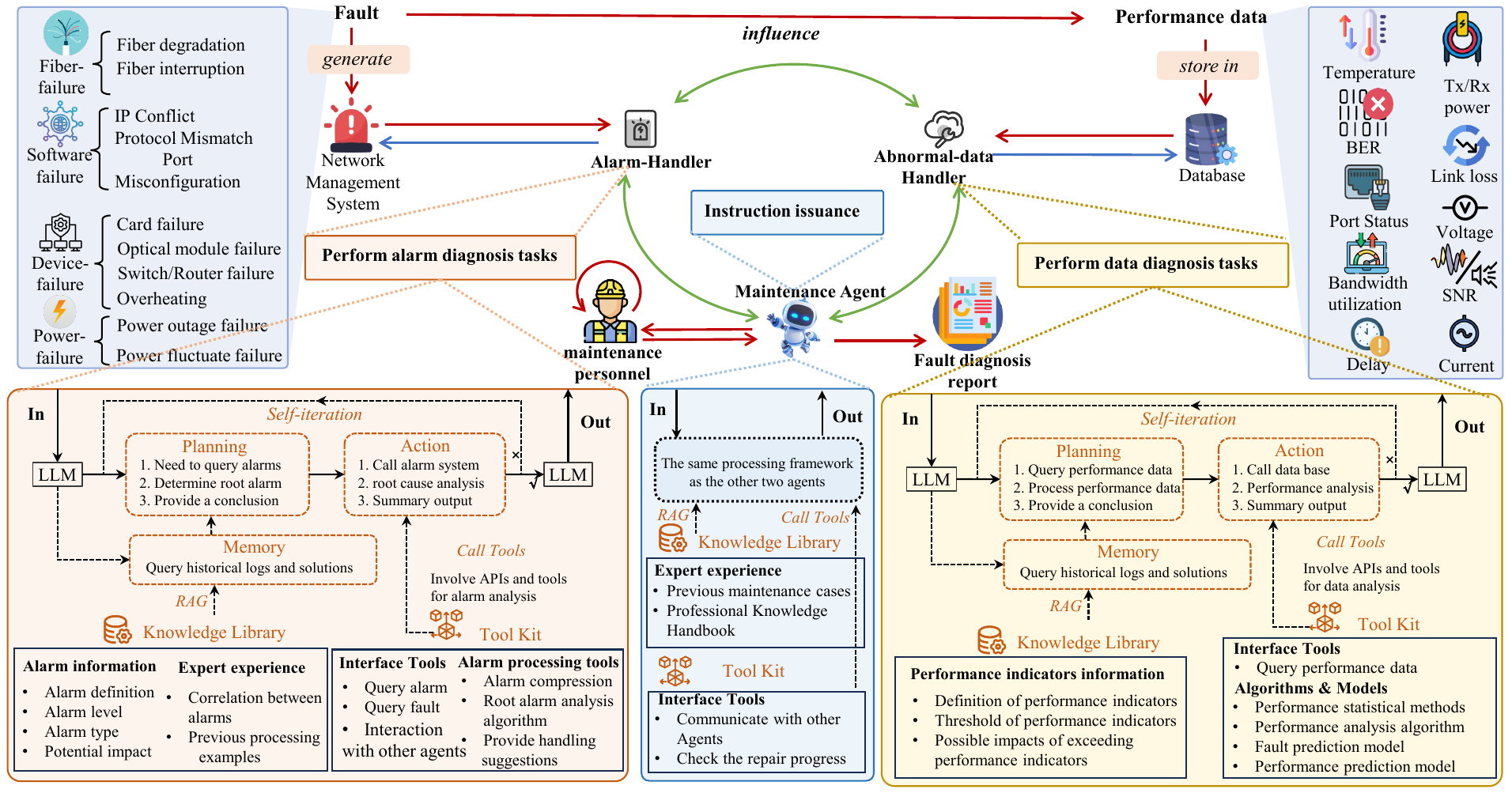}
\caption{Schematic of multi-Agent collaboration in the maintenance phase of the AONs. Faults trigger alarms and cause anomalies in performance indicators, which are detected by the Alarm Handler and Abnormal-data Handler. By correlating alarms with performance data, the Maintenance Agent identifies the fault and assists maintenance personnel in resolution. A fault diagnosis report is generated once the fault is resolved and the system returns to normal.}
\label{Fig-5.4 maintain}
\end{figure*}

During the execution, the Dynamic Operator follows the plan generated by the Operation Agent while checking the returned network state after each action. In the demonstrated trace, NMS telemetry is first retrieved to determine the current service and route state, followed by service release and provisioning requests. The resulting configuration status and predicted GSNR are then checked to verify that the modified network remains operational. Finally, the Dynamic Operator returns the execution status and updated network state to the Operation Agent for further coordination. Additionally, Dynamic Operator assists in service quality assurance, establishing traffic scheduling and priority management to ensure that network resources are allocated based on predefined service level agreement (SLA), which guarantees the real-time and emergency communications. By enforcing intelligent scheduling policies, Dynamic Operator can maintain the required quality of service (QoS) levels across diverse scenarios. Furthermore, Dynamic Operator supports software and system upgrades, ensuring smooth transitions without service disruption.

The Automated Inspector is the real-time monitoring and diagnostics Agent in optical networks, dedicated to ensuring long-term stability through continuous surveillance, predictive maintenance, and environmental awareness. Unlike the Dynamic Operator, which focuses on active optimization, the Inspector monitors key metrics such as optical power, GSNR, and bandwidth utilization to detect QoT fluctuations and anomalies. Integrated within NMS, it offers centralized visibility and automated reporting. It performs routine inspections on power flatness and spectrum usage to optimize GSNR and spectral efficiency, while also diagnosing physical infrastructure by evaluating fiber attenuation, equipment health, and identifying potential faults. Additionally, it monitors environmental conditions like temperature and humidity, enabling proactive risk mitigation and maintaining resilient network operations.

Automated Inspector can autonomously analyze anomalies and generate structured reports for further action. Detected issues are categorized and sent to the Operation Agent to determine the response strategy. If performance tuning is needed, it delegates execution to the Dynamic Operator for real-time adjustments like power tuning or path reconfiguration. The Operation Agent also periodically initiates automated inspections, during which the Inspector retrieves monitoring indicators and methodologies from its knowledge base, invoking tools and APIs for comprehensive checks. As illustrated in Fig. \ref{fig-5.3 operation2}, the Automated Inspector receives an inspection request from the Operation Agent, retrieves the required NMS performance indicators and analysis methods, and detects abnormal optical-power conditions from the returned telemetry. The action trace shows an abnormal power condition below the normal operating range. The Inspector reports these observations to the Operation Agent, which then determines whether the issue can be resolved within the current operation domain. For the demonstrated case, the Operation Agent generates a follow-up optimization request and instructs the Dynamic Operator to retrieve the relevant configuration interfaces, evaluate the optimization strategy, and apply the resulting adjustment. After configuration, the performance is re-evaluated and the updated network state is reported back to the Operation Agent.

\subsection{Maintenance phase}

\begin{figure*}[ht] 
\centering
\includegraphics[width=12cm]{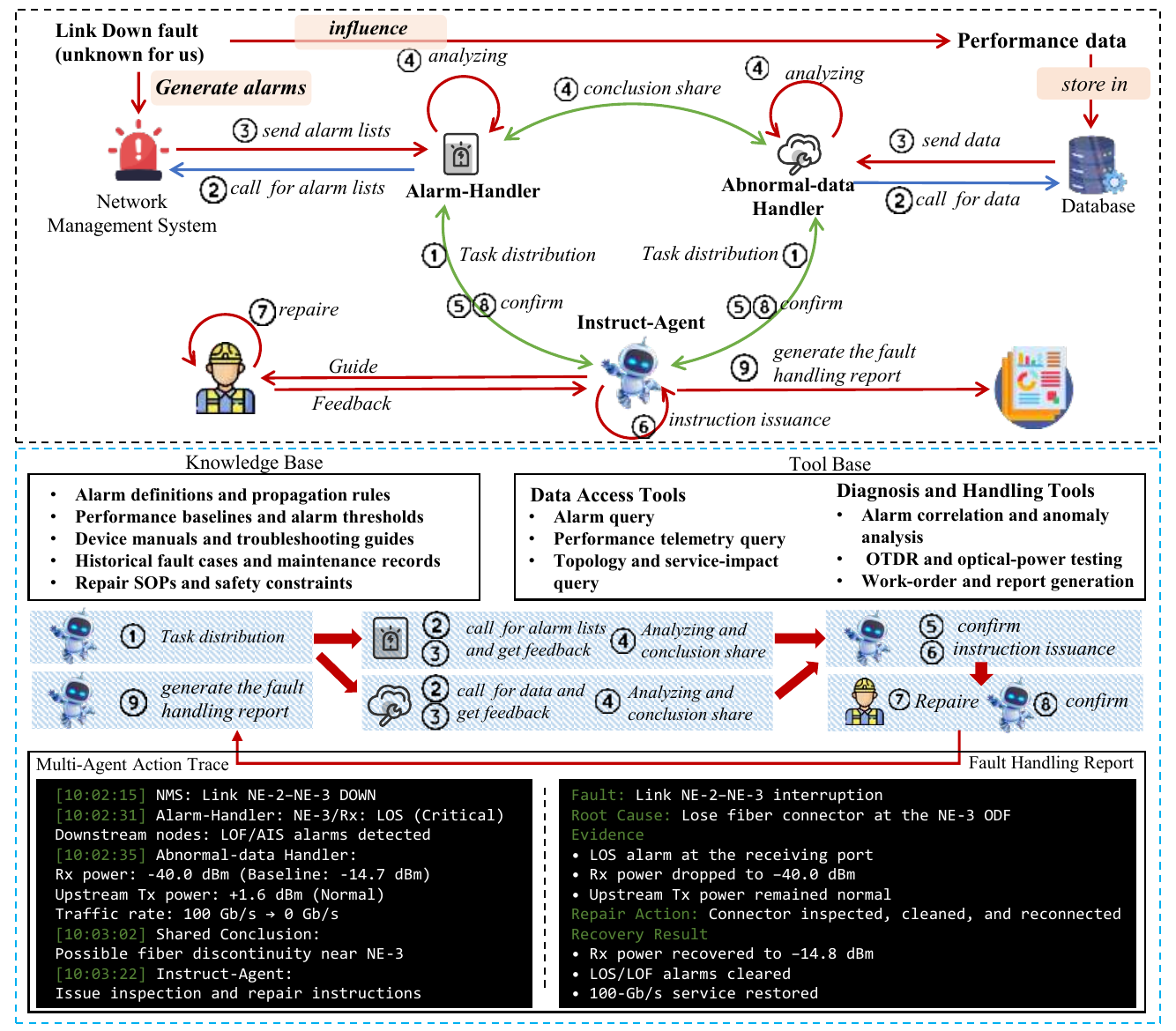}
\caption{Illustration of Agent collaboration in a specific fault diagnosis and recovery scenario during the maintenance phase. The figure demonstrates the whole process, including fault triggering, alarm and performance anomaly detection, Agent-driven root cause analysis, on-site repair guidance, and final result consolidation.}
\label{Fig-5.4 maintain2}
\end{figure*}

The goal of the maintenance phase is to quickly identify and resolve faults, thereby minimizing service interruptions and reducing economic losses as much as possible. With the continuous expansion of optical networks, the number and variety of devices and components are also growing, leading to an increasing overall complexity of the network. In the event of a failure, a large number of alarms are triggered and massive abnormal performance data are generated. Under such a complex environment, timely and accurate fault localization and repair becomes critical for ensuring optical network stability. 

Firstly, we introduce several common types of faults and the key network indicators that are typically affected when a fault occurs, as shown in Fig. \ref{Fig-5.4 maintain}. First, one of the most common optical network faults are optical fiber failures, including both gradual fiber degradation and abrupt fiber breaks. Degradation tends to occur over time and can be detected through specific performance indicators. In contrast, fiber breaks are usually sudden, often resulting from external mechanical forces or environmental changes that alter the fiber’s physical properties and rapidly degrade optical transmission. Second, the software malfunctions arise from incompatibilities between software and hardware components, leading to issues such as IP conflicts, protocol mismatches, or port misconfigurations. Third, the equipment malfunctions refer to the performance degradation of devices or modules due to factors such as aging, manufacturing defects, or environmental stress. Typical examples include failures in boards, optical modules, or overheating of network components. Fourth, the power failures frequently occur in data centers, which are caused by power outages, voltage instability, or power distribution faults. These faults often lead to abnormal or fluctuating values in key network performance indicators, directly or indirectly impacting QoS. The network indicators we monitor include temperature, voltage, current, optical power, BER, link loss, SNR, delay, port status, bandwidth, and others. Through telemetry, these indicators reflect the real-time network state and help alert maintenance personnel to potential risks. To support this, threshold-based alarms are used to alert operators when performance metrics exceed predefined limits. Although most alarms today are still threshold-driven, the concept is flexible and can incorporate multi-dimensional conditions rather than relying solely on single-metric triggers.

The overall framework of the maintenance phase and the roles of the involved Agents are depicted in Fig. \ref{Fig-5.4 maintain2}. When a fault occurs, the network management system generates alarms, while the fault may also cause abnormal changes in the performance data stored in the database. The Maintenance Agent first assigns diagnostic tasks to two specialized Agents: the Alarm Handler and the Abnormal-data Handler. The Alarm Handler retrieves and correlates alarm lists to distinguish root alarms from derived alarms. Meanwhile, the Abnormal-data Handler retrieves telemetry data and examines deviations from normal performance baselines. The two Agents exchange their intermediate conclusions and jointly narrow down the fault location and possible root cause.

These diagnostic activities are supported by a shared knowledge base and tool base. The knowledge base contains alarm definitions and propagation rules, performance baselines and alarm thresholds, device manuals and troubleshooting guides, historical fault cases and maintenance records, and repair SOPs and safety constraints. The tools are divided into two groups. Data access tools provide alarm, performance telemetry, topology, and service-impact queries. Diagnosis and handling tools support alarm correlation and anomaly analysis, OTDR and optical-power testing, and work-order and report generation.

After receiving the diagnostic results, the Maintenance Agent checks the consistency of the conclusions and determines the appropriate repair procedure based on the retrieved manuals, historical cases, and operational constraints. It then sends the fault location and repair instructions to field maintenance personnel. After the repair is completed, the Maintenance Agent coordinates with the Alarm Handler and Abnormal-data Handler to verify that the alarms have cleared and that the performance indicators have returned to their normal ranges. Finally, it consolidates the diagnostic evidence, repair actions, and recovery results into a structured fault-handling report. The conclusion-sharing links in the figure represent inter-Agent communication, which can be implemented through A2A-compatible interfaces.

\begin{figure*}[ht] 
\centering
\includegraphics[width=15cm]{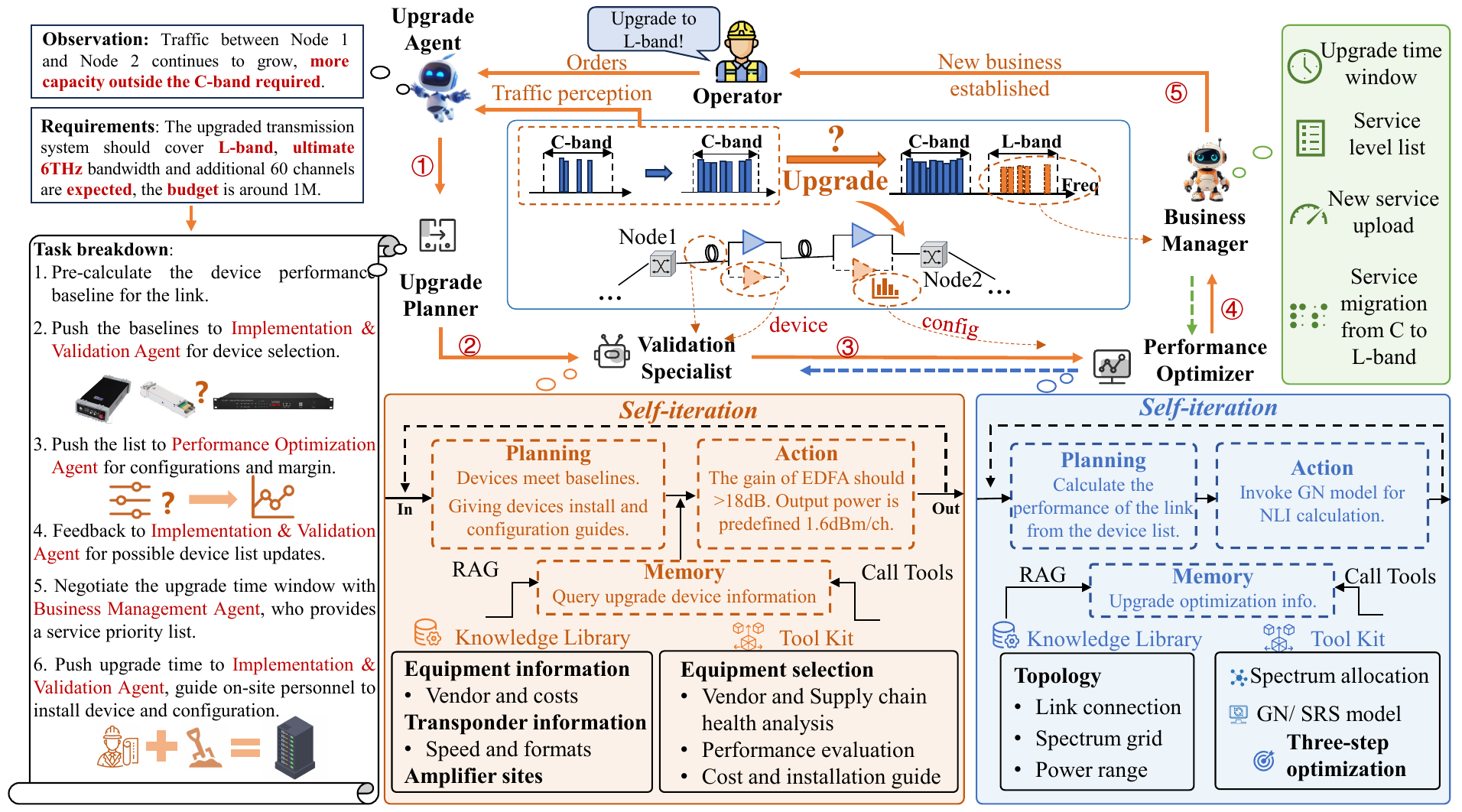}
\caption{Schematic of multi-Agent collaboration in the upgrade phase of the AONs. The figure depicts how the Upgrade Agent predicts capacity saturation and coordinates with sub-Agents, including the Upgrade Planner, Validation Specialist, Performance Optimizer, and Business Manager, to carry out a C-to-C+L band upgrade. It visualizes each Agent’s role in planning, device validation, performance optimization, and scheduling, ensuring a seamless and efficient network upgrade process.}
\label{fig-5.5 upgrade}
\end{figure*}

Next, we illustrate the fault localization and recovery process through the representative link-down case shown in Fig. \ref{Fig-5.4 maintain2}. At 10:02:15, the network management system reports that the link between NE-2 and NE-3 is down. The Maintenance Agent responds by assigning diagnostic tasks to the Alarm Handler and Abnormal-data Handler. The Alarm Handler queries the network management system and identifies a critical loss-of-signal (LOS) alarm at the receiving port of NE-3, together with downstream loss-of-frame (LOF) and alarm-indication-signal (AIS) alarms. In parallel, the Abnormal-data Handler retrieves the relevant performance data. It finds that the received optical power has dropped from a baseline of approximately -14.7 dBm to -40.0 dBm, while the upstream transmit power remains normal at approximately +1.6 dBm. The traffic rate has also decreased from 100 Gb/s to zero.

By correlating the alarm and performance evidence, the two diagnostic Agents infer that the failure is likely caused by a fiber discontinuity near NE-3. The Maintenance Agent then consults the relevant troubleshooting procedures and issues inspection and repair instructions to the field personnel. The on-site inspection confirms that a loose fiber connector at the NE-3 optical distribution frame (ODF) caused the interruption. The connector is inspected, cleaned, and reconnected. After the repair, the received optical power recovers to approximately -14.8 dBm, the LOS and LOF alarms are cleared, and the 100 Gb/s service is restored. Finally, the Maintenance Agent records the root cause, diagnostic evidence, repair action, and recovery status in a structured fault-handling report.

\subsection{Upgrade phase}

The ever-growing network traffic necessitates continuous upgrades in optical networks. During routine network management, the Upgrade Agent analyzes real-time data and historical trends to predict capacity saturation at different parts of networks. When C-band utilization approaches 80\%, the Upgrade Agent generate upgrade reports to operator, cross-referencing external context such as SLAs. Meanwhile, the operator is also monitoring such links whose capacity is approaching its limit. To prevent potential failures and accommodate new traffic, the operator can instruct the Upgrade Agent to promote this link to C+L-band transmission, as illustrated in Fig. \ref{fig-5.5 upgrade}.

\begin{figure*}[ht] 
\centering
\includegraphics[width=15cm]{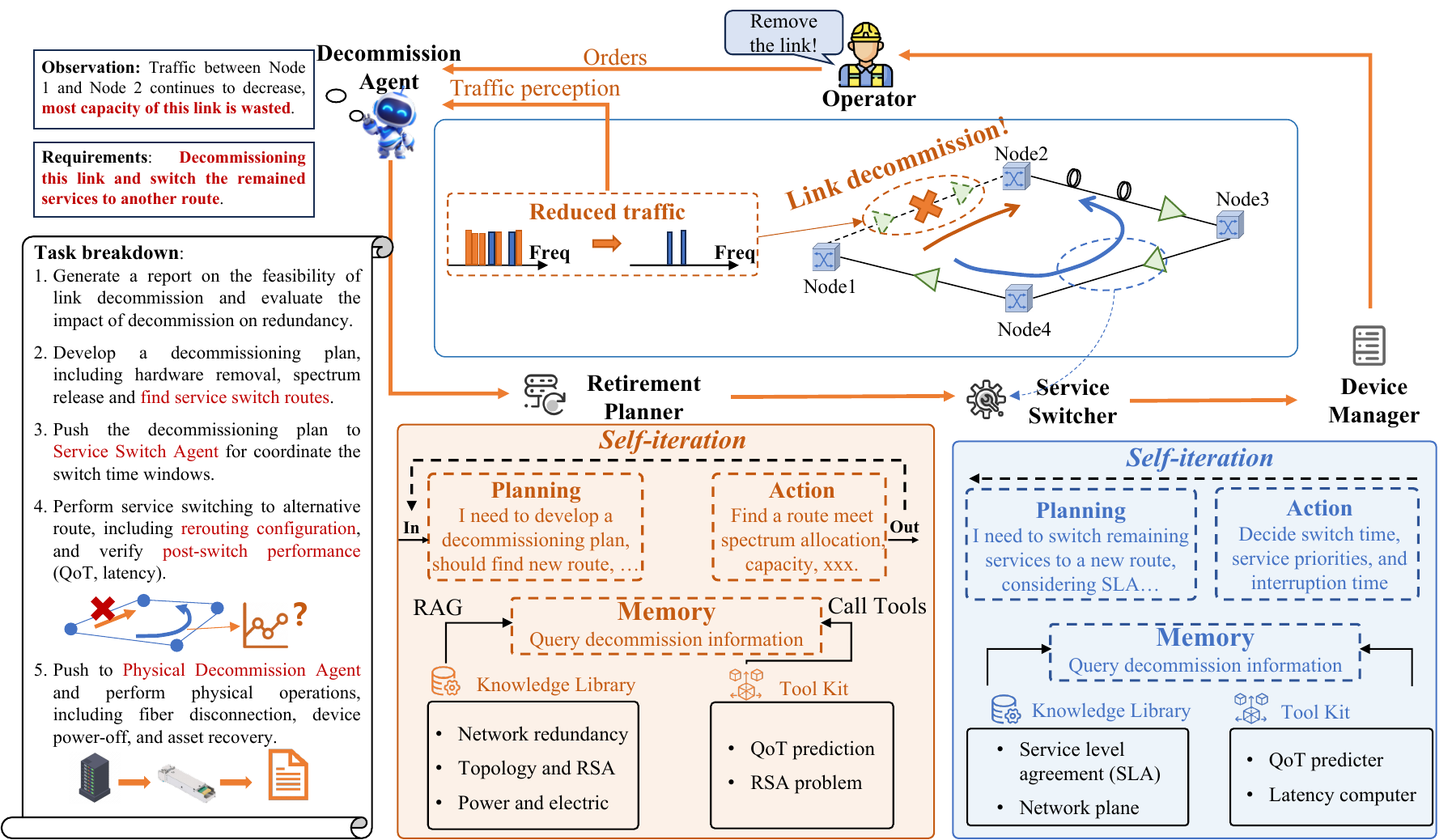 }
\caption{Schematic of multi-Agent collaboration in the decommission phase of the AONs. The figure depicts how the Decommission Agent monitors traffic and initiates decommissioning workflows when link utilization drops. It highlights the roles of the Retirement Planner, Service Switcher, and Device Manager in evaluating SLA impact, rerouting services, coordinating switching windows, and physically decommissioning network elements, ensuring minimal disruption and efficient resource reclamation.}
\label{fig-5.6 decommission}
\end{figure*}

Upon receiving the upgrade order, the Upgrade Agent summarizes the upgrade requirements, including bandwidth expansion, capacity needs, budget constraints, and other relevant factors. These requirements are then forwarded to the sub-Agent Upgrade Planner, which generates a step-by-step upgrade plan. Each step invokes specialized sub-Agents to execute specific tasks efficiently. In the upgrade from C-band to C+L-band system, the Upgrade Planner first determines the necessary devices and performance baselines for each device. For example, the output power of an L-band transponder should be around 0 dBm, while the saturation power of an L-band amplifier must exceed 23 dBm. These baselines are then passed to the Validation Specialist, which selects appropriate devices based on its infrastructure knowledge, containing records of available devices of different types, alongside vendor documentation validated against standards like ITU-T.

Within the Validation Specialist, a structured decision-making cycle is followed, consisting of planning, action, and the memory to provide necessary information. Once the required devices are selected, the Performance Optimizer is activated to calculate the optimal configurations for the upgraded L-band transmission link. This sub-Agent also follows a similar cycle, where it breaks down the task into sub-steps and formulates a CoT process to determine the best approach. Different actions are selected based on computational tools from the memory, and results are continuously analyzed for further optimization. Moreover, there is self-iteration of derived strategy from the outputs to the inputs. For example, when the L-band devices are selected, the Performance Optimizer can invoke the three-step QoT optimization algorithm \cite{song2022efficient} to find optimal configurations for the C- and L-band EDFAs. As the C-band transmission performance will also be affected by the upgraded L-band devices and traffic, fast optimization tools are necessary to mitigate transient interference during the upgrade.

After determining the optimal configurations, all relevant information is transferred to the Business Manager, which schedules an appropriate upgrade time window to minimize network disruptions. Additionally, the Business Manager ensures that critical services maintain high performance and that C-band optical performance remains within acceptable limits. It also identifies new services and existing services that need to be migrated to the L-band. Once these steps are completed, the C+L-band transmission system is successfully upgraded, ensuring seamless scalability and uninterrupted service delivery.

\subsection{Decommission phase}

With the emergence of new services, advancements in communication technologies, and shifts in communication capacity distribution across different nodes, parts of the transmission system may need to be decommissioned, or specific links with minimal services may be removed. Taking link decommissioning as an example, potential decommissioning candidates are first identified during the operational phase based on traffic and service conditions. Once a decommissioning request is initiated, the upper-level orchestration layer activates the Decommission Agent, which then continuously monitors the relevant traffic and service status within the optical network, as illustrated in Fig. \ref{fig-5.6 decommission}. For an identified candidate, the Decommission Agent evaluates the SLA and generates a comprehensive report for the operator. This report assesses the necessity and feasibility of decommissioning, considering factors such as the importance of remaining services, operational costs, and potential savings from decommissioning. Once the report is submitted, the operator makes the final decision on whether to proceed with decommissioning.

Once the decision is confirmed, the Decommission Agent forwards the decommissioning request to the Retirement Planner, which then breaks down the process into logical, feasible steps, delegating specialized tasks to sub-Agents for efficient execution. First, the Retirement Planner generates a detailed report assessing alternative routing capacity, SLA compliance, and the impact of decommissioning on network redundancy to ensure that decommissioning does not compromise service quality and that sufficient backup routes exist. It then formulates a structured decommissioning plan covering hardware removal/retention, spectrum release, and service rerouting. Following a structured cycle, planning, action, and self-iteration, the plan is further broken down into sub-tasks using CoT reasoning. The Agent identifies suitable alternative service routes using network knowledge and computational tools, ensuring these routes provide adequate capacity, available spectrum slots, and compliance with latency constraints.

Once the plan is developed, it is pushed to the Service Switcher for implementation. The Service Switcher coordinates the switching time window to ensure minimal network disruption, reroutes services to alternative paths, and conducts post-switch performance validation, including QoT and latency checks. After successful service switching, the process moves to the Device Manager, which executes the physical decommissioning operations, including fiber disconnection, device power-off, and asset recovery management, such as optical module removal and inventory registration. After the retirement actions and post-decommissioning validation are completed, the Decommission Agent concludes the active decommissioning process and becomes inactive until another decommissioning task is initiated. Upon completion of these steps, the link decommissioning process is finalized, ensuring efficient resource optimization while maintaining network stability.

\section{Outlook}

Despite the promising potential of the multi-Agent technique in AON, there are still challenges that needs to be addressed before the full implementation in practice. 

\subsection{Hallucination risks}

Hallucination \cite{ye2023cognitive, huang2025survey} refers to cases where LLMs generate fluent and coherent outputs that are factually incorrect, logically inconsistent, or contextually irrelevant. Hallucinations may arise from insufficient domain knowledge during pre-training, noisy data correlations, or inference-stage factors such as ambiguous prompts and limited context windows \cite{adlakha2024evaluating}. This issue poses a major challenge for the reliable deployment of LLM Agents in optical networks, where high accuracy and faithful consistency are essential. In optical networks, hallucinations often manifest in the form of incorrect parameter suggestions, nonexistent device models, or logically flawed configuration schemes \cite{wang2025wireless}. Such errors are particularly critical in tasks such as QoT estimation, failure localization, and control plane automation, where precision and consistency are paramount. Mitigating hallucinations requires domain-aware validation and system-level safeguards rather than relying solely on model improvements \cite{tonmoy2024comprehensive}. In the proposed framework, generated results are first supervised within the corresponding lifecycle division. The Division Agent evaluates the outputs of its subordinate AI Experts against task objectives, network constraints, and expected quality, and triggers refinement and re-execution when necessary. This local closed loop continues until an acceptable result is obtained or a predefined iteration limit is reached, after which unresolved tasks are escalated to the upper level or human operators. For actions that may affect the physical network, the Security Manager provides an additional safety gate by checking generated configurations or commands against operational and security constraints. Generated decisions should be verified against network state, topology, and operational policies before execution. In particular, DTON can serve as a validation sandbox to evaluate generated strategies and configurations against physical-layer constraints and real-time network conditions before deployment. High-risk or service-affecting actions can further require human-in-the-loop confirmation before execution. After deployment, network states and service performance are continuously monitored, and corrective actions or rollback to a previously validated configuration can be initiated if unacceptable outcomes are detected. Besides, hallucination mitigation should be regarded as a joint responsibility of both LLM Agents and deterministic network infrastructures to ensure trustworthy operation.

\subsection{Model selection and deployment}

In practical deployment, the hierarchical multi-Agent framework requires that Agents can be adapted to the complexity, latency requirement, data sensitivity, and computational demand of individual tasks. High-level Agents, for example, the Network Director, are responsible for global objective interpretation, cross-domain information integration, long-horizon planning, task decomposition, and coordination among multiple Agents. Cloud infrastructure can host large models and shared knowledge services for computationally intensive tasks. In contrast, Agents for operation, monitoring, security, and other latency-sensitive tasks can be smaller language models, deployed at the edge or locally at network sites, particularly when rapid responses or local processing of operational data are required. A hierarchical escalation mechanism can also be adopted, where a lightweight local model handles routine requests and invokes a more capable remote model only when the task is ambiguous, complex, or high-risk. Moreover, considering that some tasks require faster, more accurate, and complex calculations, we design and recommend using workflows with tools instead of calling LLM to accelerate task implementation and ensure accuracy.

\subsection{Token costs}

Token length constraints present a fundamental limitation for applying LLMs in optical networks, where decision-making often relies on large volumes of structured and unstructured data. Practical tasks may require the model to simultaneously process multi-span transmission parameters, per-channel configurations, network topologies, historical monitoring records, and real-time telemetry data, leading to rapidly increasing context size and token consumption as network scale grows. This challenge becomes more severe in full LCM tasks, where decisions frequently depend on information accumulated across multiple stages. Addressing this issue requires efficient context management \cite{zhang2025survey}, including hierarchical information abstraction, skill-based function calling, retrieval-based context selection, and memory-augmented architectures. 

In our proposed multi-Agent framework, this issue can be mitigated through the dedicated Memory Container, which serves as an intermediate layer for context compression and management. By leveraging efficient token compression, structured information encoding, and selective retrieval mechanisms, the Memory Container can distill large-scale network data into compact and task-relevant representations, thereby reducing token consumption while maintaining essential global context. The context requirements are also closely coupled with model size and computational resource allocation.

\subsection{Long-horizon memory}

Long-horizon memory and stateful reasoning remain critical challenges for applying LLMs to optical network LCM, where decisions spans extended time horizons, and depend on evolving network states, historical configurations, and accumulated performance data. However, conventional LLMs operate in a largely stateless manner, relying only on the current context window and lacking persistent awareness of prior interactions or long-term system evolution. This limitation is particularly evident in sequential tasks such as fault diagnosis, dynamic reconfiguration, and upgrade planning, where earlier actions directly influence subsequent decisions. Addressing this challenge requires external memory and state management mechanisms, including structured repositories, memory-augmented architectures, and DT-assisted synchronization. In our framework, the Memory Container further acts as a persistent state management module that supports storage, update, and retrieval of network state information across different lifecycle stages. By maintaining evolving network states, it enables temporally consistent and context-aware reasoning over long operational horizons.

In addition, persistent network knowledge and short-term task contexts can be managed separately so that only the information required for the current reasoning step is provided to the LLM. The context requirements are also closely coupled with model size and computational resource allocation, with more demanding or high-priority tasks receiving greater computing resources and lightweight tasks being handled by smaller models or edge resources. Such coordinated management of context, model scale, and computing resources can reduce context and computational costs while maintaining the responsiveness and reasoning capability required for optical network lifecycle management.

\section{Conclusion}

This paper has systematically envisioned the LLM Agent-driven evolution of optical networks toward AONs. First, the technical foundations of LLM and Agent are illustrated, with particular emphasis on domain adaptation techniques and performance evaluation considerations in optical networking scenarios. From a system-level perspective, this work provided a structured decomposition of the optical network lifecycle into six key phases, and analyzed the associated tasks, data requirements, and operational constraints in each stage. Based on these insights, we presented a hierarchical multi-Agent framework that organizes heterogeneous Agents across different functional layers, enabling coordinated perception, reasoning, and action. This framework offers a unified perspective for integrating LLM-driven intelligence into optical network management, bridging the gap between high-level decision-making and low-level operational execution.

Beyond architectural design, an important takeaway of this work lies in highlighting the role of LLM Agent as a potential enabler of closed-loop, adaptive, and context-aware control in optical networks. By coupling LLM reasoning capabilities with domain knowledge, real-time data, and external tools, Agent systems can move toward more flexible and scalable management strategies compared to conventional pipeline-based solutions. In evolution of optical networks toward AONs, this work aims to present a structured and forward-looking perspective on leveraging LLM Agent for the autonomous LCM. Rather than offering a definitive solution, it seeks to outline a conceptual architecture and provide technical insights for researchers and developers to explore the evolving landscape of AONs. It is expected that continued efforts in model development, system integration, and real-world validation will be necessary to fully realize the potential of AONs.

\section*{Funding}National Natural Science Foundation of China (62522104 and U24B20133).

\bibliography{reference}


\end{document}